\UseRawInputEncoding 
\documentclass[journal]{IEEEtran}
\usepackage{mathrsfs}
\usepackage[T1]{fontenc}
\usepackage{cite}
\usepackage{psfrag}
\usepackage{subfigure}
\usepackage{url}
\usepackage{stfloats}
\usepackage{array}
\usepackage{amsthm}
\usepackage{booktabs}  
\usepackage{algorithm} 
\usepackage{setspace}
\usepackage{algorithmic} 
\usepackage{amsmath,amsfonts,amsbsy,bm,paralist,theorem,ifthen,color}
\usepackage{graphicx}
\usepackage{placeins}
\usepackage{float}
\usepackage{hyperref}
\usepackage{tabularx,colortbl}
\usepackage{multirow}
\usepackage{multicol}
\usepackage{arydshln}
\usepackage{colortbl}
\usepackage{tipa}
\usepackage{siunitx}
\usepackage{amssymb}

\ifCLASSINFOpdf
\fi
\begin{document}

\title{Artificial Intelligence Enabled Radio Propagation for Communications---Part I: Channel Characterization and Antenna-Channel Optimization}

\author{Chen Huang,~\IEEEmembership{Member,~IEEE,}
        Ruisi~He,~\IEEEmembership{Senior~Member,~IEEE,}
        Bo~Ai,~\IEEEmembership{Senior~Member,~IEEE,}
        Andreas~F.~Molisch,~\IEEEmembership{Fellow,~IEEE,}
        Buon~Kiong~Lau,~\IEEEmembership{Senior~Member,~IEEE,}
        Katsuyuki~Haneda,~\IEEEmembership{Member,~IEEE,}
        Bo~Liu,~\IEEEmembership{Senior~Member,~IEEE,}
        Cheng-Xiang~Wang,~\IEEEmembership{Fellow,~IEEE,}
        Mi~Yang,~\IEEEmembership{Member,~IEEE,}
        Claude~Oestges,~\IEEEmembership{Fellow,~IEEE,}
        Zhangdui~Zhong,~\IEEEmembership{Senior~Member,~IEEE}
       \\ (\emph{Invited Paper})
       \vspace{-1 cm}
\thanks{This work is supported by National Key R\&D Program of China under Grant 2020YFB1806903, the National Natural Science Foundation of China under Grant 61922012, U1834210, and 61961130391, the State Key Laboratory of Rail Traffic Control and Safety under Grant RCS2020ZT010, and the Fundamental research funds for the central universities under Grant 2020JBZD005 and I20JB0200030. (\emph{Corresponding authors: Ruisi He; Bo Ai.})}
\thanks{C. Huang is with the Purple Mountain Laboratories, Nanjing, 211111, China, and also with the National Mobile Communications Research Laboratory, School of Information Science and Engineering, Southeast University, Nanjing, 210096, China (email: huangchen@pmlabs.com.cn). Part of this work was done when he was a Ph.D candidate at Beijing Jiaotong University.}
\thanks{R. He, M. Yang, and Z. Zhong are with the State Key Laboratory of Rail Traffic Control and Safety, and the Key Laboratory of Railway Industry of Broadband Mobile Information Communications, Beijing Jiaotong University, Beijing 100044, China (email: {ruisi.he, myang, zhdzhong}@bjtu.edu.cn).}
\thanks{B. Ai is with the State Key Lab of Rail Traffic Control and Safety, Beijing Jiaotong University, Beijing 100044, China, with the School of Information Engineering, Zhengzhou University, Zhengzhou 450001, China, and also with Peng Cheng Laboratory, Shenzhen 518055, China (e-mail: aibo@ieee.org).}
\thanks{A. F. Molisch is with the Ming Hsieh Department of Electrical and Computer Engineering, University of Southern California, Los Angeles, CA 90089 (email: molisch@usc.edu).}
\thanks{B. K. Lau is with the Department of Electrical and Information Technology, Lund University, Lund 22100, Sweden (email: bklau@ieee.org).}
\thanks{K. Haneda is with the Department of Radio Science and Engineering, Aalto University, 00076 Aalto, Finland (email: katsuyuki.haneda@aalto.fi).}
\thanks{B. Liu is with the School of Engineering, University of Glasgow, Glasgow G12 8QQ, U.K. (email: Bo.Liu@glasgow.ac.uk).}
\thanks{C.-X. Wang is with the National Mobile Communications Research Laboratory, School of Information Science
and Engineering, Southeast University, Nanjing, 210096, China, and also with the Purple Mountain Laboratories, Nanjing, 211111, China (email: chxwang@seu.edu.cn).}
\thanks{C. Oestges is with the Institute of Information and Communication Technologies, Electronics and Applied Mathematics, Universite Catholique de Louvain, 1348 Louvain-la-Neuve, Belgium (email: claude.oestges@uclouvain.be).}

}


\maketitle

\begin{abstract}
To provide higher data rates, as well as better coverage, cost efficiency, security, adaptability, and scalability, the 5G and beyond 5G networks are developed with various artificial intelligence techniques. In this two-part paper, we investigate the application of artificial intelligence (AI) and in particular machine learning (ML) to the study of wireless propagation channels. It firstly provides a comprehensive overview of ML for channel characterization and ML-based antenna-channel optimization in this first part, and then it gives a state-of-the-art literature review of channel scenario identification and channel modeling in Part II. Fundamental results and key concepts of ML for communication networks are presented, and widely used ML methods for channel data processing, propagation channel estimation, and characterization are analyzed and compared. A discussion of challenges and future research directions for ML-enabled next generation networks of the topics covered in this part rounds off the paper.
\end{abstract}


\IEEEpeerreviewmaketitle

      \vspace{-1 cm}
\section{Introduction}

\IEEEPARstart{T}{he} dramatic increase of the numbers of wireless users and wireless applications brings new demand and challenges for wireless communication networks. The 5G and beyond 5G (B5G) networks are expected to provide higher data rates, as well as better coverage, cost efficiency, security, adaptability, and scalability \cite{wang2020artificial,tataria20216g}. Since 2020, 5G communication has begun to be deployed worldwide \cite{wang20206g, 7999256}, whereas studies of 6G wireless communication networks have started in academic and industrial research labs to further enhance MBB, expand the application and coverage of the internet of things (IoT), and make networks/devices more intelligent. These new application scenarios give 6G network a series of new performance requirements: 10--100 million devices connections with the peak data rate of 1-10 TB/s; the mobility that needs to be supported rises to higher than 1000 km/h to accommodate ultra-high-speed train (uHST), unmanned aerial vehicle (UAV) \cite{khuwaja2018survey}, and satellites; latencies need to be reduced to fractions of 1 ms to account for tactile internet and other real-time control applications; and reliability of 5 or even 7 times has to be achieved for mission-critical applications. Also, to provide global coverage, 6G wireless networks will expand from terrestrial communication networks to space-air-ground-sea integrate networks.

The study of propagation channels is a fundamental aspect of any wireless communication system design, network optimization, and performance evaluation \cite{fan2018wireless, 9362269, 686758, 4052596, 6280628}. Therefore, to realize 6G networks to meet the requirements above, the corresponding wireless channels need to be thoroughly studied. However,  the massive---in terms of number of devices, number of antennas, bandwidth, etc.---scenarios not only pose a challenge in performing dedicated measurement campaigns, but it also leads to massive amounts of data that need to be processed and analyzed \cite{575640}. Classical techniques for such analysis, e.g., parameter estimation, tracking, clustering, and characterization, are generally less suited for such large amounts of data, either because of the resulting overhead, or because they might miss important relationships within the data.

On the other hand, artificial intelligence (AI) has been developed to ``simulate the human intelligence processes by machines, especially computer systems''~\cite{bi2015wireless}. Machine learning (ML) is a branch of AI that enables machines to learn from a massive amount of data and make decisions and/or perform actions accordingly without been given any specific commands. With the help of continually increasing computing power, ML techniques have achieved great success in big data processing for many applications, e.g., image processing, natural language processing, and data mining. Consequently, ML techniques have also been widely applied to various problems in communications networks, and are expected to be an integral part of next-generation communication networks~\cite{chen2019artificial,bin2020machine, songyan2020learn, lin2020adaptive, MBOOK, 8952905}.

\subsection{Development of AI and ML}AI techniques have made rapid advances in many domains since the last decade, including communication and electronic engineering. An elaborate definition of AI is ``a system's ability to correctly interpret external data, to learn from such data, and to use those learnings to achieve specific goals and tasks through flexible adaptation'' \cite{haenlein2019brief}. Hence, AI researchers aim to build intelligent agents to achieve this goal.

ML is a subset of AI, which studies algorithms that allow computer programs to automatically improve through experience \cite{dietterich1997machine}. ML can be generally classified into supervised learning, unsupervised learning, and reinforcement learning~\cite{dietterich1997machine}. Deep learning (DL) \cite{goodfellow2016deep}, is a subset of ML and has attracted tremendous attention in recent years. The major difference compared to the traditional ML is the way of using training data. Besides, artificial neural networks (ANN) are the backbone of DL algorithms, whereas for traditional ML, the learning machines vary and are not limited to ANN. Widely used deep neural network (DNN) structures include Convolution Neural Network (CNN), Restricted Boltzmann Machine, Long Short-Term Memory (LSTM), etc \cite{shrestha2019review}. Due to the no-free-lunch theorem, DL requires much more data and computing resources than the traditional ML.

Knowing the concepts of AI and ML, a natural question is how AI and ML techniques can support research on antennas and propagation. The answer is that practical problems must first be formulated in a particular mathematical way, such that AI/ML techniques can serve as solvers. Note that AI/ML techniques are not the only way to solve these problems, but research works show clear advantages of them compared to the conventional methods, at least under certain circumstances. The mathematical problems that AI/ML techniques focus on are described as i) \emph{regression}, which identifies the relationships between a dependent variable (i.e., output data) and one or more independent variables (i.e., input data); ii) \emph{classification}, which uses a set of training data for which the feature and category membership or label is known to identify to which of a set of categories a new instance belongs to; iii) \emph{clustering}, which naturally groups a set of objects with the aim that objects in the same group are more similar to each other compared to those in other groups, without any training data.

\begin{table*}[]
\caption{Summary of ML-based Applications for Communications.}\label{tb_overview}
\center
\begin{tabular}{|l|l|l|l|}
\hline
\multicolumn{2}{|l|}{Category}                                             & Typical algorithms                    & Applications                                                                                                                                                                 \\ \hline
\multirow{6}{*}{Regression}      & \multirow{3}{*}{Supervised learning}    & Support vector machine                & \multirow{6}{*}{\begin{tabular}[c]{@{}l@{}}$\bullet$ Channel parameter estimation\\ $\bullet$ Channel characterization/modeling\\$\bullet$  Channel prediction\\ $\bullet$ Antenna system optimization\end{tabular}} \\ \cdashline{3-3}[0.8pt/2pt]
                                 &                                         & Relevance vector machine              &                                                                                                                                                                              \\ \cdashline{3-3}[0.9pt/2pt]
                                 &                                         & Aritificial neural network            &                                                                                                                                                                              \\ \cdashline{2-3}[0.9pt/1pt]
                                 & Un-supervised learning                  & Bayesian learning                     &                                                                                                                                                                              \\ \cdashline{2-3}[0.9pt/1pt]
                                 & Reinforcement learning                  & Boltzmann exploration algorithm       &                                                                                                                                                                              \\ \cdashline{2-3}[0.9pt/1pt]
                                 & Meta-learning                           & Deep transfer-based meta-learning     &                                                                                                                                                                              \\ \hline
\multirow{10}{*}{Classification*} & \multirow{7}{*}{Supervised learning}    & Support vector machine                & \multirow{10}{*}{\begin{tabular}[c]{@{}l@{}}$\bullet$ LoS/NLoS identification\\ $\bullet$ Scenario identification\\ $\bullet$ Antenna selection optimization\\ $\bullet$ Cluster member optimization\end{tabular}}   \\ \cdashline{3-3}[0.9pt/2pt]
                                 &                                         & Relevance vector machine              &                                                                                                                                                                              \\ \cdashline{3-3}[0.9pt/2pt]
                                 &                                         & Decision trees                        &                                                                                                                                                                              \\ \cdashline{3-3}[0.9pt/2pt]
                                 &                                         & Random forest                         &                                                                                                                                                                              \\ \cdashline{3-3}[0.9pt/2pt]
                                 &                                         & Aritificial neural network            &                                                                                                                                                                              \\ \cdashline{3-3}[0.9pt/2pt]
                                 &                                         & K-Nearest neighbor                    &                                                                                                                                                                              \\
                                 \cdashline{3-3}[0.9pt/2pt]
                                 &                                         & Naive Bayesian                        &                                                                                                                                                                              \\
                                 \cdashline{2-3}[0.9pt/1pt]
                                 & \multirow{2}{*}{Un-supervised learning} & Variational Bayesian                  &                                                                                                                                                                              \\ \cdashline{3-3}[0.9pt/2pt]
                                 &                                         & Hypothesis test                       &                                                                                                                                                                              \\ \cdashline{2-3}[0.9pt/1pt]
                                 & Reinforcement learning                  & Monte Carlo tree search               &                                                                                                                                                                              \\ \hline
\multirow{8}{*}{Clustering*}      & \multirow{2}{*}{Supervised learning}    & Hidden Markov model                   & \multirow{8}{*}{\begin{tabular}[c]{@{}l@{}}$\bullet$ MPC cluster identification\\ $\bullet$ LoS/NLoS identification\end{tabular}}                                                                \\ \cdashline{3-3}[0.9pt/2pt]
                                 &                                         & K-Nearest neighbor                    &                                                                                                                                                                              \\ \cdashline{2-3}[0.9pt/1pt]
                                 & \multirow{6}{*}{Un-supervised learning} & KpowerMeans/K-Means                   &                                                                                                                                                                              \\ \cdashline{3-3}[0.9pt/2pt]
                                 &                                         & Density-based spatial clustering      &                                                                                                                                                                              \\ \cdashline{3-3}[0.9pt/2pt]
                                 &                                         & Region competition                    &                                                                                                                                                                              \\ \cdashline{3-3}[0.9pt/2pt]
                                 &                                         & Hough-transform-based clustering      &                                                                                                                                                                              \\ \cdashline{3-3}[0.9pt/2pt]
                                 &                                         & Kernel-power-density-based clustering &                                                                                                                                                                              \\ \cdashline{3-3}[0.9pt/2pt]
                                 &                                         & Fuzzy-C-Means                         &                                                                                                                                                                              \\ \hline
\end{tabular}
\\  {\footnotesize{*The algorithms are classified according to the applications in communications rather than typical ML categories. Typical clustering ML methods are}}
\leftline {\footnotesize{usually un-supervised, but there are also supervised learning methods adopted for the MPC clustering problem.}}
\vspace{- 0.8 cm}
\end{table*}

\subsection{Application of ML to Propagation Channels}
AI techniques have been introduced into communications over the last two decades. They have addressed many bottlenecks that the conventional methods are not able to resolve, from communication system design to propagation channel research~\cite{he2021wireless, 9395374, 9158524, 8952905, 9212600}, with the latter being the focus of this section\footnote{In this part, the focus is on channel characterization and antenna-channel optimization. In Part II \cite{PartII}, we review the research on ML-based scenario identification and channel modeling.}.

Channel feature extraction is a necessary step for channel analysis and modeling~\cite{6060881, 9104014, 5979196, 299557, 8852816}. Some basic channel parameters, such as channel gain and Doppler power spectrum, can be calculated directly from the channel transfer function. However, for some complex channel characteristics, such as power angular spectrum (PAS) and multipath component (MPC) parameters, a targeted channel parameter estimation algorithm is needed. For multipath channel models, such as tapped delay line (TDL), WINNER II, and cluster-based models, it is necessary to obtain joint parameters such as power, delay, and angle of the MPCs. For conventional parameter estimation, the main challenges they face are complexity traps and the applicability of diverse scenarios. The low-energy and low-computing devices in the Internet of Things are likely to be difficult to carry more complex conventional methods. Therefore, the ML-based method may provide unexpected and insightful new solutions. Based on analyses from a large number of measurement campaigns, the obtained MPCs are usually distributed in groups, as know as clusters. As early as 1972, the cluster structure of MPCs was characterized in the delay domain in \cite{turin1972statistical}. The classical Saleh-Valenzuela model developed in \cite{Saleh1986ASM} provides a general and very widely used cluster-based channel model, which describes both inter-cluster and intra-cluster characteristics. This model was generalized in \cite{spencer2000modeling} to include both delay and angular domains, whereas the cluster structure of MPCs in the geometrical map based on the physical environment is revealed in \cite{Fuhl1998UnifiedCM}, which can also be mapped to a delay/angle description. Since then, more and more channel models and standards are developed based on a cluster structure, e.g., COST 259 \cite{Molisch2006TheCD, Asplund2006TheC2}, COST 2100 \cite{Liu2012TheC2},  3GPP Spatial Model \cite{3gpp}, and WINNER \cite{Meinila2009WINNERIC, KystiWINNERIC}.

Given that the human brain is good at pattern identification, the MPC clusters can be identified by human inspection, as in \cite{Karedal2007AMS, Cramer2002EvaluationOA, Czink2007ClusterCI, Chong2005AGS}. However, human inspection is subjective, not repeatable, and may be unreliable. Moreover, due to the development of propagation measurement techniques, the amount of measurement data has increased dramatically, making human inspection impractical. On the other hand, ML methods have been found to have a great advantage in processing multi-dimension data \cite{Chen2019ArtificialNN, Huang2019MachineLearningBasedDP}, which naturally correspond to the channel measurement data in multiple domains, e.g., delay, and angular domains. Specifically, data clustering methods are one of the hot topics in ML \cite{Jain1999DataCA} and have become a powerful tool for MPC cluster identification \cite{he2018clustering}. In addition, analyzing time-varying channels requires the capturing of the evolving behavior and lifetime of the MPCs/clusters. In this case, the time-varying channel characterization requires not only clustering but also tracking the MPCs in consecutive snapshots, and ML methods has been proved to have great ability on recognizing data pattern which naturally fits the requirement of MPC/clustering tracking.

Although propagation channels can be characterized and modeled independently of antennas, by de-embedding the antenna properties such as radiation pattern, polarization, and impedance matching, the performance of communication systems depends on the overall physical channel, which includes antenna effects \cite{fan2013emulating, fan2013antenna, 6863647, 7875483, 8691774}. Specifically, the interaction between an antenna and the propagation channel determines the channel as seen by the transmitting and receiving antenna ports. Moreover, since the propagation channel properties are largely determined by the surrounding environment, the antenna represents an opportunity (as a spatial filter) for optimizing the antenna-channel interaction from the viewpoint of performance for the application of interest, be it communication, localization, or otherwise. For illustration, the received power will increase if the gain of the receiving antenna gain increased in the direction(s) of the incoming signal, assuming no polarization mismatch between the channel and the antenna. In reality, the optimization is complex given a varying multipath propagation channel with different properties in magnitude, phase, polarization, delay, angle of arrival/departure, as well as an increased degree of freedom in the design of receiving antennas with the use of multiple elements for array/MIMO operation.
Hence, optimizing antenna-channel interaction is another important subject for optimizing wireless system performance.

ML is useful to address this optimization problem due to its ability to create mathematical models for the complex problem involving myriads of antenna and channel parameters. Importantly, useful models can be obtained even with only limited availability of training data. Different from the use of ML to speed up parametric optimization in classical antenna design by reducing the number of required full-wave simulations to obtain detailed antenna properties, the focus of this survey is on optimizing the far-field properties of antennas for improved interaction with the propagation channel and thus enhanced system performance. Specifically, the conventional antenna optimization algorithms, e.g., \cite{heath2001antenna, molisch2004mimo}, are usually computationally expensive, and highly dependent on the quality and completeness of the channel knowledge. On the contrary, the ML-based antenna optimization usually employs an offline training process, e.g., \cite{joung2016machine, he2018transmit, cai2019antenna, zhang2020efficient, chen2019intelligent}. After training process, the trained method can be implemented online with low computation complexity, with greatly reduce the latency of the network.

Table \ref{tb_overview} summarizes the commonly used ML methods for communication applications that are investigated in this two-part paper. It is noteworthy that some tracking/matching algorithms, which are not usually considered as ML methods, are also implemented with ML methods for time-varying cluster identification, dynamic channel parameter estimation, etc. They are also investigated in this paper but not included in Table \ref{tb_overview}.

This paper is organized as follows. Section II summarizes the ML-based channel feature extraction techniques, which include channel parameter estimation, MPC tracking, and MPC clustering, whereas the ML-based design and optimization of antenna-channel interactions are covered in Section III. Section IV provides a discussion of  challenges and possible future research avenues for the topics above. Section V concludes the paper, while scenario identification, channel modeling and prediction, and the discussion of these two open topics will be handled in Part II \cite{PartII}.

\begin{table*}[]
\caption{Summary of Channel Parameter Estimation.}\label{tb_channelextraction}
\center
\begin{tabular}{|l|l|l|c|}
\hline
Category                                                  & Typical Algorithm & Parameter Type                                 & Existing Works                          \\ \hline
\multicolumn{1}{|c|}{\multirow{5}{*}{Traditional method}} & $\bullet$ Bartlett          & \multirow{2}{*}{PAS}                           & \cite{bartlett1948smoothing}     \\ \cdashline{2-2}[0.8pt/2pt]\cdashline{4-4}[0.8pt/2pt]
\multicolumn{1}{|c|}{}                                    & $\bullet$ MUSIC             &                                                & \cite{schmidt1986multiple}       \\ \cdashline{2-4}[0.8pt/2pt]
\multicolumn{1}{|c|}{}                                    & $\bullet$ ESPRIT            & Signal angle                                   & \cite{roy1986estimation}         \\ \cdashline{2-4}[0.8pt/2pt]
\multicolumn{1}{|c|}{}                                    & $\bullet$ SAGE              & MPC joint parameters (delay, angle, amplitude) & \cite{fleury1999channel}         \\ \cdashline{2-4}[0.8pt/2pt]
\multicolumn{1}{|c|}{}                                    & $\bullet$ RiMAX             & MPC joint parameters; DMC                      & \cite{thoma2004multidimensional} \\ \hline
\multirow{5}{*}{ML-based method}                          & $\bullet$ SVM               & Path loss                                      & \cite{uccellari2017application}  \\ \cdashline{2-4}[0.8pt/2pt]
                                                          & $\bullet$ ANN               & Channel excess attenuation                     & \cite{bai2019prediction}         \\ \cdashline{2-4}[0.8pt/2pt]
                                                          & $\bullet$ SVM-PCA           & AoA and ASA                                    & \cite{yang2021machine}           \\ \cdashline{2-4}[0.8pt/2pt]
                                                          & $\bullet$ Bayesian learning & \multirow{2}{*}{DoA}                           & \cite{liu2012efficient}          \\ \cdashline{2-2}[0.8pt/2pt] \cdashline{4-4}[0.8pt/2pt]
                                                          & $\bullet$ RVM               &                                                & \cite{cal2013relevance}          \\ \cdashline{2-4}[0.8pt/2pt]
                                                         & $\bullet$ EKF        &    MPC joint parameters; DMC        & \cite{4732313} \\
                                                          \hline
\end{tabular}
     \vspace{-0.3 cm}
\end{table*}

\section{Channel Feature Extraction}
\subsection{Channel Parameter Estimation}

The PAS quantifies the signal energy at different angles. It thus reflects the spatial characteristics of the channel and plays an important role in directional channel modeling. Traditional channel angle parameter estimation algorithms rely on antenna array signal processing, such as Bartlett beamforming~\cite{bartlett1948smoothing}, and similarly delay characteristics are obtained from Fourier transformation of the transfer function. Conversely, high-resolution parameter estimation (HRPE) overcomes the resolution limitation inherent to classical techniques like the Fourier or beamforming methods and shows excellent performance in multi-dimensional MPC parameter estimation. These techniques can be divided into subspace-based methods and parametric estimators. A prominent representative for the former one is Multiple Signal Classification (MUSIC), which decomposes the signal space into signal subspace and noise subspace and then determines the direction of the source by the eigenvalue of covariance matrix \cite{schmidt1986multiple}. Estimation Signal Parameter via Rotational Invariance Techniques (ESPRIT) solves the angle information of signals based on the rotation invariance principle of sub-arrays \cite{roy1986estimation}. Among the parametric estimators, iterative maximum-likelihood estimators such as Space-Alternating Generalized Expectation-maximization (SAGE) and RiMAX are widely used. The basic idea of the SAGE algorithm is to decompose a multidimensional joint estimation problem into multiple one-dimensional maximum-likelihood estimation problems. When estimating the parameters for a certain MPC, SAGE is to fix a sweep parameter to only one of the parameters of a single MPC while others in the same MPC and other MPCs are not swept \cite{fleury1999channel}. RiMAX uses a similar basic idea as the SAGE algorithm, but adds dense multipath components (DMCs) to the signal model, and uses a different way of iteration, to jointly estimate the specular reflection path and DMCs of channels \cite{thoma2004multidimensional}.

ML methods, which have been widely developed in recent decades, provide effective solutions for many nonlinear mapping problems for which it is difficult to establish explicit mathematical expressions. For channel parameter estimation solutions based on ML, the basic idea is to take a large amount of input data, and then obtain a regression model through a learning algorithm, so that the outputs of the model are channel parameters. Ref. \cite{huang2018big} introduces the prediction and modeling of channel parameters driven by ML, and points out applications of ML algorithms in the estimation of channel parameters such as path loss, angle information, and channel impulse response (CIR). The application of DL to HRPE is discussed in \cite{huang2018deep}, whereas an ML-based angle-of-arrival (AoA) estimation approach is proposed in \cite{yang2021machine}. In the proposed method, an estimation model is obtained by using a support vector machine (SVM) based on a large number of actual array measurement data. Then, the obtained model is used to realize AoA estimation according to the channel snapshots collected by the antenna array. Extended Kalman Filter (EKF) is used in \cite{4732313} to track multipaths over time while estimating the multipath parameters at a snapshot through maximum likelihood.

Overall, novel ML-based channel parameter extraction solutions have broad application prospects. At the same time, the propagation channels for emerging wireless communication systems face an evolutionary trend to a higher frequency, diversified and complex scenarios, and more abundant devices. In this context, parameter extraction algorithms also face some challenges:
\subsubsection{Performance in a non-stationary environment.} Vehicular, high-speed railway and satellite communications are gradually brought into the service range of wireless communication systems, and these high-speed mobile devices result in fast-time-varying as well as strongly non-stationary channels (i.e., not only the channel realizations but also the channel {\em statistics} change within a short time). Therefore, the real-time ability and accuracy of the ML-based parameter extraction algorithm in non-stationary channels have to be verified.

\subsubsection{Selection of input and output parameters.} ML-based parameter estimation algorithms essentially establish the relationship between input data and target outputs, so the selection of inputs is extremely important. Existing input features mainly include environmental information, signal strength, and channel impulse response. In addition, the requirements of system design and channel modeling are increasingly diversified, which drives the demand for parameter estimation, such as super-resolution estimation of multipath information, real-time estimation of scenario category and angle information, etc. Therefore, it is necessary to investigate and optimize reasonable parameter selection schemes for different requirements.

\subsubsection{Acquisition of big data of measurement data.} ML-based parameter extraction algorithms need a huge amount of real channel measurement data. Therefore, the design of channel-sounder \cite{fan2018flexible} and channel measurement campaign will play an important role in future ML-based algorithms. However, it is expensive and time-consuming to carry out channel measurement in various environments, which will become the bottleneck of novel algorithms - this is especially true for mmWave and THz bands, and high-mobility scenarios such as aircrafts and trains.

Therefore, while evolving the channel measurement methods, it is necessary to explore an efficient way of data utilization, such as using an AI-driven denoising algorithm to improve the signal-to-noise ratio of data, and obtaining massive data through the combination of measurement and simulation. A DOA estimator based on a spatially over-complete array output formulation is proposed in \cite{liu2012efficient}. Numerical results show that the proposed method has excellent performance, especially in demanding scenarios such as low signal-to-noise ratio (SNR). A relevance vector machine (RVM)-based detection algorithm of MPC arrival times and complex amplitudes is proposed in \cite{cal2013relevance}, which can be used to filter MPCs of simulated power delay profiles embedded with noise. Another widely used training method for ML algorithms - not only for channel parameter extraction but also other wireless applications - is based on stochastic channel models such as Quadriga \cite{Jaeckel14TAP} or ray tracers such as Wireless InSite \cite{Remcom}.
Table \ref{tb_channelextraction} summarizes the widely used channel parameter estimation methods and the parameter type of each method.

\subsection{MPCs' Cluster Identification}
\label{ClusterIdentification}

As briefly introduced in the introduction, the clustering technique is one of the important applications of ML, which naturally meets the requirement of grouping the MPCs with similar channel characteristics for further modeling. On the other hand, for ML-based channel modeling, the MPCs' cluster identification is also an important procedure as pre-processing to help ML-method learn the data pattern more efficiently. According to the clustering methodology, the existing cluster identification algorithm can be categorized into the following types.

\subsubsection{Shape-Based Cluster Identification}
The cluster structure of MPCs is firstly revealed in the power-delay domain. The MPCs belonging to the same cluster can be modeled as having a PDP that follows a single exponential decay  \cite{Saleh1986ASM}. Inspired by this, shape-based cluster identification has been proposed to recognize the MPC clusters in a PDP. The key idea is to use a fitting method to see if the shape of the envelope of the MPCs matches a particular distribution. By adjusting the cluster members, i.e., MPC, to seek the best fitting result, the MPC cluster can be well identified.
Following this idea, \cite{Shutin2004ClusterAO} trains a Hidden Markov Model (HMM) to learn the distribution of the MPCs in the PDP and optimizes the cluster members by using the Viterbi algorithm \cite{Rabiner1989ATO}.
Instead of seeking the single exponential decay distribution, another solution is to express the PDP on a semi-logarithmic scale (for power domain as in $y$-axis) so that exponential decay profiles will be displayed as straight lines with constant slopes, e.g., \cite{Chuang2007AutomatedIO, Corrigan2009AutomaticUC}. By seeking straight lines that best fit the MPCs, the proposed algorithm is able to identify the clusters efficiently and objectively. Specifically, Ref. \cite{Corrigan2009AutomaticUC} builds an observation window to separate the big cluster. By applying a threshold of the slope, the proposed algorithm can improve the clustering accuracy on small clusters.
Different from the ideas above, kurtosis is adopted in \cite{Gentile2013UsingTK} to measure the shape of the MPC distributions and applies the Region Competition method \cite{Zhu1996RegionCU} to divide the MPCs into different clusters. The kurtosis is commonly used to measure the peakedness of a certain distribution, whereas the Region Competition method is a widely used group member determination algorithm in computer vision.

Overall, the strong point of the shape-based cluster identification is that the clustering procedure doesn't require much prior knowledge, e.g., the number of clusters. Besides, it only focuses on the MPC distribution in PDP, which leads to a low computation complexity because of relatively low dimensionality. Nevertheless, the identification accuracy is also limited due to a lack of angular information during the clustering.

\subsubsection{Optimization-Based Cluster Identification}
For any data clustering method, noise impacts the clustering accuracy significantly. As introduced before, there is always some unremovable noise effect during the MPC parameter estimation, which may also have an influence on the MPC cluster identification. In this case, optimizing the channel measurement data (raw data) by removing the noise or reconstructing the CIR is one way to improve the MPC cluster identification accuracy. Following this idea, optimization-based cluster identification takes into account the behaviors of MPCs in CIRs during the clustering to improve the performance. A sparsity-based clustering method is proposed in \cite{He2016OnTC, He2016ASC}, which aims to recover the ideal CIRs (without the noise) by solving a sparsity-based optimization problem that incorporates the characteristics of the CIRs, then identify the cluster using the recovered CIRs. Gaussian Mixture Model (GMM) is introduced in \cite{Li2020ClusteringAI, Li2018ClusteringIW} to recognize the clusters in the angle-delay domain, where the Expectation-Maximization algorithm \cite{Tzikas2008TheVA} is adopted to seek the initial parameters of GMM and the Variational Bayesian (VB) algorithm \cite{Ma2011BayesianEO} is applied to further optimize the number of Gaussian distributions.

\begin{figure*}[!t]
    \centering
    \includegraphics[width=0.99 \textwidth]{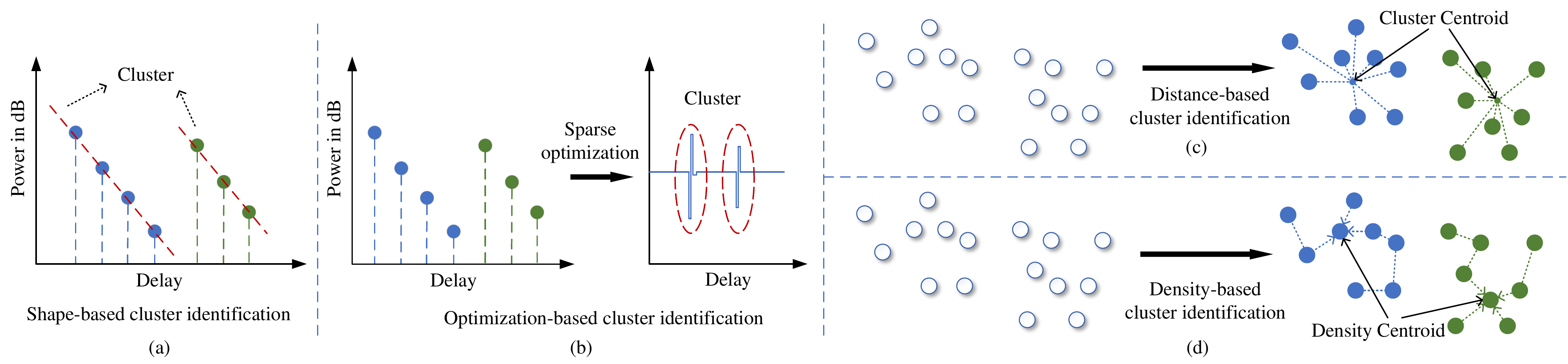}
     \caption{The comparison of the key idea of (a) the shape-based;(b) the optimization-based;(c) the distance-based; and (d) the density-based cluster identification.}
    \label{fig_clustering1}
     \vspace{-0.2 cm}
    \end{figure*}

Both the shape-based and the optimization-based cluster identification achieve good performance in terms of identifying the clusters in PDP and both require little prior knowledge for identification, whereas the optimization-based method can be further extended to introduce the angular characteristic to improve the cluster identification accuracy. It is noteworthy that, the shape-based identification can also extend to the angular domain by assuming a particular shape of cluster PAS. Nevertheless, it is difficult to determine a general/specific PAS shape for the shape matching procedure\footnote{To the best of the authors' knowledge, the existing shape-based cluster identification methods mostly fall into the delay-power domain.}. The difference between the key idea of the shape-based and the optimization-based method is illustrated in Figs. \ref{fig_clustering1}(a) and (b).

\subsubsection{Distance-Based Cluster Identification}
The very nature of the data clustering is to group the elements (MPCs) based on their similarity to each other. Consequently, many studies identify the MPC clusters based on the similarity/distance among all MPC's parameters. There are several ways to measure the similarity between different MPCs, the widely used measures include i) \emph{Squared Euclidean Distance (SED)}; ii) \emph{Normalized Euclidean Distance (NED)}; and iii) \emph{Multipath Components Distance (MCD)}. Both the SED and the NED are widely used in data clustering problems; the SED focuses on the natural difference between each parameter, whereas the NED focuses on the ratio difference. By using the SED, a sequential clustering-based algorithm is proposed in \cite{Salo2005AutomaticCO} to identify the MPC clusters. However, to properly measure the similarity of the MPCs in the different domains, e.g., delay and angular domain, it is necessary to compare the parameters on the same scale, in other words, the parameters in different domains need to be normalized first. Therefore, \cite{Steinbauer2002HowTQ} proposes the MCD, which normalizes the delay and angle parameters in their respective domains before calculating the distance measures. The cluster identification performance of using the SED and the MCD is compared in \cite{Czink2006ImprovingCP}, where the MCD achieves better performance for measuring the similarity of MPCs. Since then, many studies have developed the MPC cluster identification based on the MCD. Ref. \cite{Czink2005AutomaticCO} uses the MCD within the hierarchical tree clustering algorithm to identify the MPC cluster. A K-Power-Means (KPM) algorithm is proposed in \cite{Czink2006AFF}  based on the classic K-means algorithm \cite{1017616}, which iteratively groups the MPCs based on the power-weighted MCD between each MPC and each cluster centroid. Ref. \cite{mota2011clustering} further improves the KPM by initializing the cluster centroid with the MPC having the highest power. Instead of using K-means, which is a hard-dividing method, a fuzzy-C-means (FCM)-based MPC clustering algorithm, which is a soft-dividing method, is proposed in \cite{Schneider2009ClusteringOM}.

From the propagation point of view, each MPC cluster in the channel may correspond to a physical reflection object. Since it is the interaction objects in the environment that contributes to producing MPCs, the corresponding clusters usually have different propagation distances (delay) and different angles of arrival/departure. Therefore, it is possible to distinguish the MPC clusters based on the signal arriving time and the signal arriving angle. Follow this idea, \cite{Samimi20153DSC, Samimi20163DMS} propose time-cluster-spatial lobes, which measure the difference between the propagation time and the angular spread interval to distinguish the clusters with a certain threshold.

Benefiting from the great power of the data clustering algorithm, the distance-based cluster identification generally shows high identification accuracy with acceptable computational complexity. Moreover, by using the MCD, the distance-based method can well identify the clusters in multiple domains, i.e., power-delay-angular domain. However, most of the existing distance-based methods require prior knowledge of, e.g., the number of clusters or the initial position of the cluster centroid. Thus, the identification accuracy is sensitive to the initial parameters for the algorithm. However, the best initial parameters settings highly depend on the processed data, so there is usually no ``golden rule'' for this.

\subsubsection{Density-Based Cluster Identification}
Major physical reasons causing the cluster structure of MPCs are the fine structure of reflecting objects (e.g., windowsills and other protrusions on house walls, the rough surface of the reflecting object, diffraction at/around object edges). In this case, the MPCs near the cluster centroid usually have a higher density compared to the edge of the cluster. Hence, some studies identify the clusters based on this distribution property. The density-based spatial clustering for application with noise (DBSCAN) \cite{Ester1996ADA} algorithm is proposed in \cite{Gan2015ClusterLC}  to identify the MPC clusters. Based on this, a kernel-power-density-based clustering (KPD) algorithm is developed in \cite{He2017AnAC, He2017AKA}, which defines a kernel density to incorporate the modeled behavior of MPCs and identifies the clusters based on the local density of each MPC, whereas the power of the MPCs is taken into account as the weight factor for the density.

Similar to the distance-based cluster identification, density-based clustering can well identify the MPC clusters in multiple domains since the MPC density is independently calculated in the delay domain and angular domain, e.g., \cite{Czink2006AFF} and \cite{He2017AKA}; the main difference of the key idea between them is given in Figs. \ref{fig_clustering1}(c) and (d). Furthermore, the density-based method does not require prior knowledge like the number of clusters or the initial position of cluster centroids. Thus, the influence of the initial parameter settings on clustering performance is well controlled.

\begin{table*}[]
\caption{Summary of the Use of Channel Features for Cluster Identification.}\label{tb_channelclustering}
\center
\begin{tabular}{|l|c|l|l|}
\hline
Category                           &  Data domain     & Common tools & Existing works      \\
\hline
\multirow{3}{*}{Shape-based}            & PDP                    & $\bullet$ HMM                                  & \cite{Shutin2004ClusterAO},                                                       \\ \cdashline{2-4}[0.8pt/2pt]
                                        & PDP                    & $\bullet$ Least   squares                      & \cite{Chuang2007AutomatedIO, Corrigan2009AutomaticUC}                           \\ \cdashline{2-4}[0.8pt/2pt]
                                        & PDP                    & $\bullet$ Kurtosis measure                     & \cite{Gentile2013UsingTK}                                                         \\ \hline
\multirow{2}{*}{Optimization-based}     & PDP                    & $\bullet$ Sparse optimization                  & \cite{He2016OnTC, He2016ASC}                                                      \\ \cdashline{2-4}[0.8pt/2pt]
                                        & Angle-delay   domain   & $\bullet$ GMM+VB                               & \cite{Li2020ClusteringAI, Li2018ClusteringIW}                                   \\ \hline
\multirow{8}{*}{Distance-based}         & Angle-delay domain     & $\bullet$ Sequential clustering                & \cite{Salo2005AutomaticCO, Czink2006ImprovingCP}                                \\ \cdashline{2-4}[0.8pt/2pt]
                                        & Angle-delay   domain   & $\bullet$ Hierarchical tree clustering         & \cite{Czink2005AutomaticCO}                                                       \\ \cdashline{2-4}[0.8pt/2pt]
                                        & Angle-delay   domain   & $\bullet$ KPM                                  & \cite{Czink2006AFF}                                                               \\  \cdashline{2-4}[0.8pt/2pt]
                                        & Angle-delay   domain   & $\bullet$ Enhanced KPM                         & \cite{mota2011clustering}                                                       \\ \cdashline{2-4}[0.8pt/2pt]
                                        & Angle-delay   domain   & $\bullet$ FCM                                  & \cite{Schneider2009ClusteringOM}                                                  \\ \cdashline{2-4}[0.8pt/2pt]
                                        & Angle-delay   domain   & $\bullet$ KM                                   & \cite{Huang2017ANP, Huang2017ANT}                                                 \\ \cdashline{2-4}[0.8pt/2pt]
                                        & Angle-delay   domain   & $\bullet$ Data stream                          & \cite{Wang2017AFO}                                                                \\  \cdashline{2-4}[0.8pt/2pt]
                                        & Angle-delay  domain   & $\bullet$ Fixed parameter interval threshold                          & \cite{Samimi20153DSC, Samimi20163DMS}                                                              \\ \hline
\multirow{2}{*}{Density-based}          & Angle-delay domain     & $\bullet$ DBSCAN                               & \cite{Gan2015ClusterLC}                                                           \\ \cdashline{2-4}[0.8pt/2pt]
                                        & Angle-delay   domain   & $\bullet$ KPD                                  & \cite{He2017AnAC, He2017AKA}                                                      \\ \hline
\multirow{3}{*}{Vision-based } & PDP/Angle-delay domain & $\bullet$ Visual inspection                    & \cite{Karedal2007AMS,   Cramer2002EvaluationOA, Czink2007ClusterCI, Chong2005AGS} \\ \cdashline{2-4}[0.8pt/2pt]
                                        & Angle-delay domain     & $\bullet$ Hough transform                      & \cite{Cai2018HoughTransformBasedCI}                                               \\ \cdashline{2-4}[0.8pt/2pt]
                                        & Angle-delay  domain   & $\bullet$ PASCT                                & \cite{Huang2019APB, Huang2018ANT, huang2018analysis, Wu2020ANP}                                    \\ \hline
Evolving-based                          & Angle-delay domain     & $\bullet$ Trajectory-based                     & \cite{Huang2020TrajectoryJointCA,   Huang2020ClusteringPE}                        \\ \hline
\end{tabular}
\vspace{-0.3 cm}
\end{table*}

\subsubsection{Computer-Vision-Based Cluster Identification}
Before automatic cluster identification methods were devised, MPC clusters were identified by human inspection as in~\cite{toeltsch2002statistical, Cramer2002EvaluationOA, karedal2004uwb,Chong2005AGS,Czink2007ClusterCI}. Despite the somewhat subjective operation model in human-eyeball clustering, there are still some basic criteria that are generally applied, including i) the shape of the potential cluster; ii) the distribution pattern of the MPCs' delay and angle; and iii) the power distribution of all MPCs. All these principles are visual-based, hence, it is also possible to use an image-processing method to recognize the MPC cluster. Inspired by this, some studies focus on computer-vision-based cluster identification. With the consideration of the delay behaviors of the MPCs, a Hough-Transform-based clustering algorithm is present in \cite{Cai2018HoughTransformBasedCI} for V2V channels, which exploits the Hough-Transform \cite{Illingworth1988ASO} to recognize the trajectory of MPC in the delay domain and merges the recognized trajectory into clusters. A PAS-based clustering and tracking (PASCT) algorithm without any high-resolution parameter extraction is proposed in \cite{Huang2019APB, Huang2018ANT, huang2018analysis}; it introduces the maximum-between-class-variance method \cite{Otsu1979ATS} to separate the potential cluster groups from the background noise and further divides the clusters by using the density-peak-search method \cite{Pedersen2016ClusteringBF}. The PASCT algorithm identifies the clusters directly from the PAS, which can be fast obtained by applying the Bartlett-beamformer \cite{Babtlett1948SmoothingPF}. A similar method is also adopted in \cite{Wu2020ANP}, where the cluster is recognized from the PAS by using image denoising, coarse-grained segmentation, and fine-grained segmentation.

The vision-based cluster identification follows an intuitive approach and thus can provide identification results that conform to human observation and benefit from the rapid development of computer vision science.

\subsubsection{Evolution-Based Cluster Identification}
   \begin{figure}[!t]
    \centering
    \includegraphics[width=0.40 \textwidth]{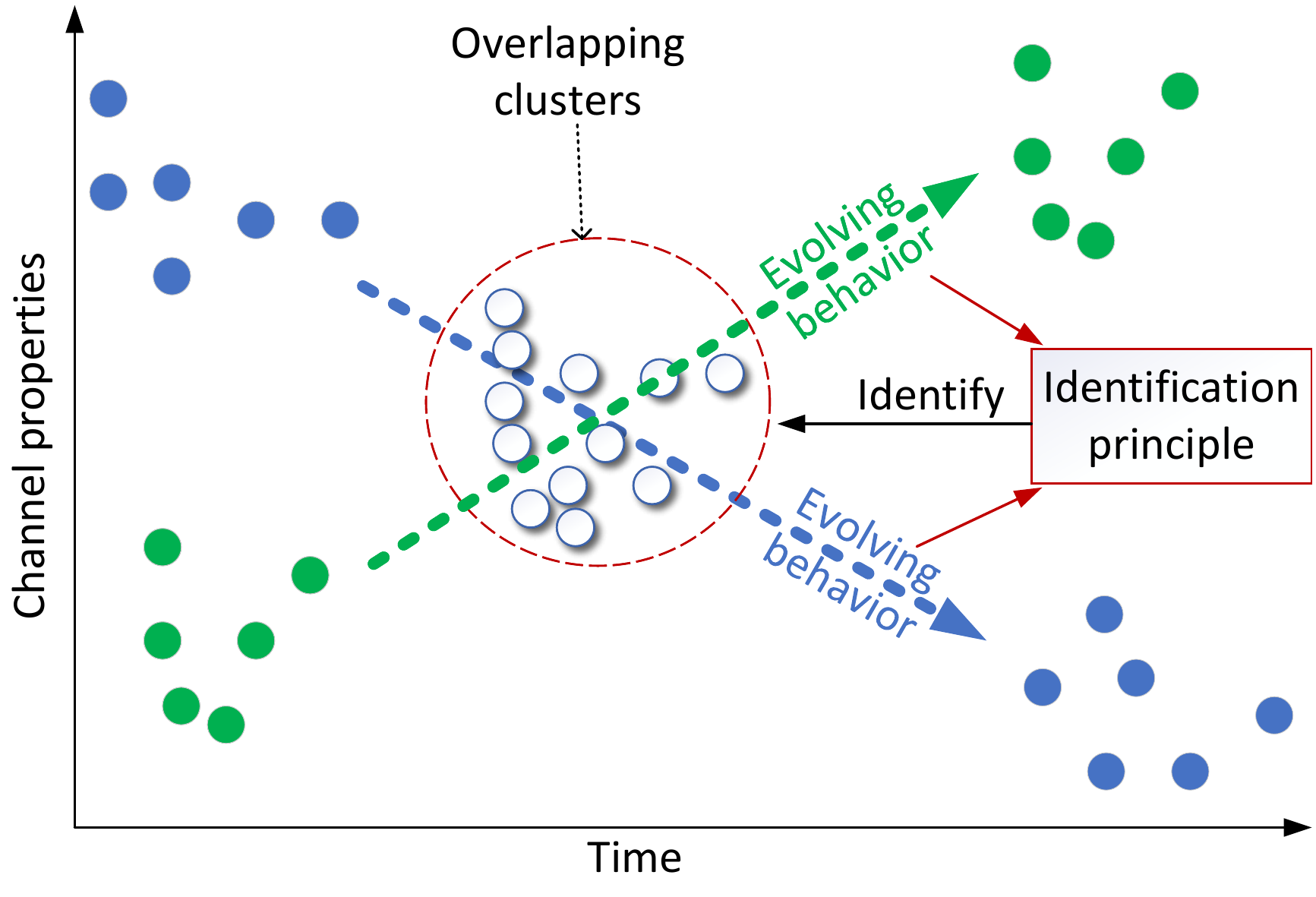}
    \caption{Illustration of the evolving-based cluster identification.}
    \label{fig_evo_tracking}
         \vspace{-0.5 cm}
    \end{figure}

The methods introduced above only consider the MPCs' behavior in a single snapshot, i.e., distribution behavior, though the evolving behavior (dynamic change) of the MPCs is another important characteristic for identifying the clusters in time-varying channels. Theoretically, a certain cluster shall correspond to a specific reflection object, except LoS.\footnote{It has been found the LoS signal also shows a cluster structure in many measurements \cite{huang2020geometry, yang2020measurements}.} In this case, the dynamic changes of the transmitter, receiver, and reflection objects lead to an evolving behavior of the MPCs, such that MPCs belonging to a particular cluster show the same/similar evolving behavior, which is distinct from the evolution of other clusters. This is a crucial feature to identify clusters that overlap each other for a period of time, but have notably different evolving patterns, as shown in Fig. \ref{fig_evo_tracking}. Following this idea, a moving probability-based clustering algorithm is proposed by using the Kuhn-Munkres (KM) algorithm in~\cite{Huang2017ANP, Huang2017ANT} for the time-varying channels, whereas \cite{Wang2017AFO} identifies the dynamic clusters based on the data-stream clustering method~\cite{Aggarwal2003AFF}. An evolving similarity is defined in \cite{Huang2020TrajectoryJointCA, Huang2020ClusteringPE} based on~\emph{the similarity of the shape of MPC's trajectories} and\emph{ the distance between the MPC}. Based on the evolving similarity, the KPD \cite{He2017AKA} algorithm is applied to identify the dynamic clusters.

The evolving-based cluster identification takes advantage of the nature of MPCs arising from the physical environment feature and the long-term evolving behavior of the clusters; it can thus provide better identification accuracy than clustering the MPCs in each snapshot independently. Meanwhile, how to capture the evolving behavior is a critical point for the evolving-based method, which is further introduced in Section~\ref{cluster_tracking}.

In summary, the ML-based cluster identification solutions show significant ability to identify the MPC clusters for further analyzing and modeling, and each identification solution has its own advantages and limitations. Considering there is no identification ground truth of the measurement data, how to properly quantify the performance of the identification method is still a challenging issue. One way is by evaluating the identification method by using synthetic data generated from a cluster-based channel model, where the identification ground-truth of the synthetic data can be easily acquired. Another solution is to use the statistical figure-of-merits for data clustering algorithms, e.g., the Calinski-Harabasz index \cite{Maulik2002PerformanceEO},  Generalized Dunn's (GD) index \cite{Pakhira2004ValidityIF}, Xie-Beni (XB) index \cite{Wang2007OnFC}, or Davies-Bouldin index \cite{Maulik2002PerformanceEO}. Several measuring methods including the indices above are compared in \cite{Mota2013EstimationOT}, where the XB index and GD index generally show good performance for evaluating the MPC cluster identification result. Table. \ref{tb_channelclustering} summarize the existing clustering methods for wireless channels.

\subsection{Multipath Component/Cluster Tracking}
\label{cluster_tracking}
Analyzing time-varying channels requires the capturing of the evolving behavior and lifetime of the MPCs/clusters. In this case, the time-varying channel characterization requires not only clustering but also tracking the MPCs in consecutive snapshots. To this aim, some ML methods, e.g., the Kalman filter \cite{Brown1983IntroductionTR}, the Hungarian
algorithm \cite{Bazaraa1977LinearPA}, and the Kuhn-Munkres algorithm \cite{Munkres1957AlgorithmsFT}, are used for MPC/cluster tracking mission.

In the past, there were mainly two types of MPC/cluster tracking: i) track the MPCs from snapshot to snapshot first, and then identify the cluster trajectory based on its MPC member; and ii) identify the cluster in each snapshot first and then track the cluster centroid. From a mathematical point of view, these two types can be further categorized into the following approaches.

\subsubsection{Distance-Based Tracking}
   \begin{figure}[!t]
    \centering
    \includegraphics[width=0.42 \textwidth]{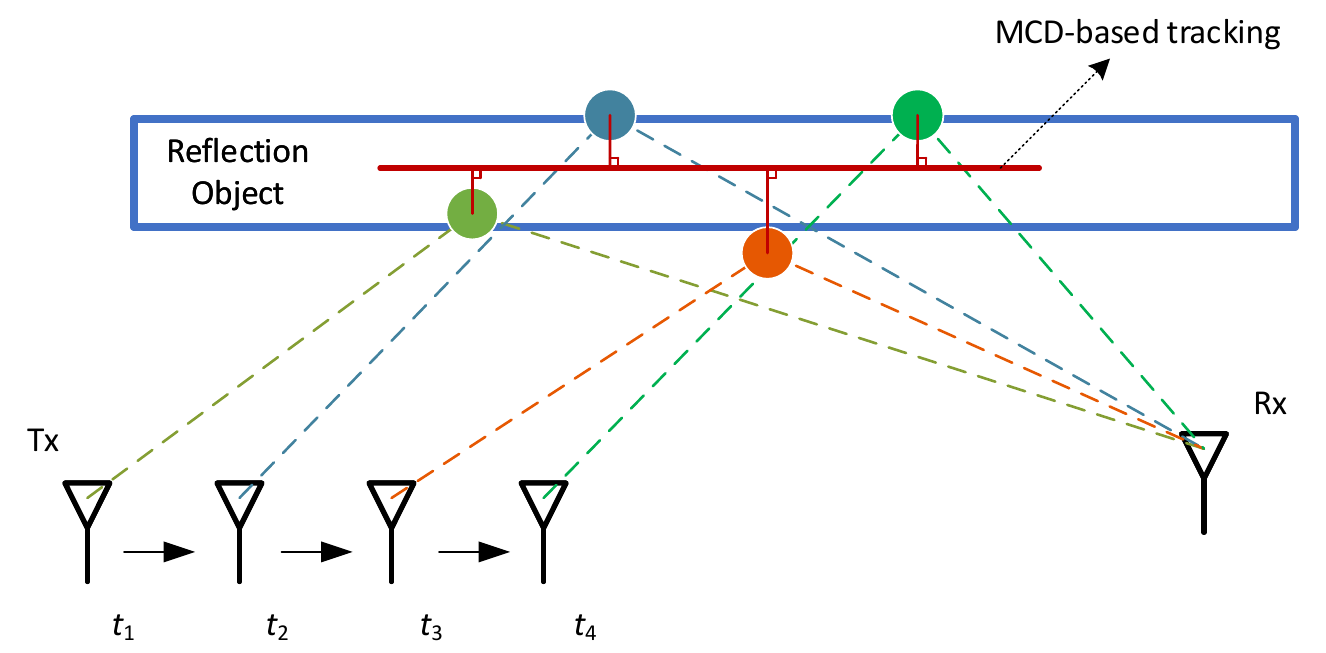}
    \caption{Key idea of tracking the reflection point along the surface in \cite{Hanpinitsak2017MultipathCA}.}
    \label{fig_tracking_dis}
         \vspace{-0.5 cm}
    \end{figure}
For time-varying channels, the dynamic MPCs/clusters are usually captured at consecutive times (snapshots). It is widely accepted that the variation of the MPCs/clusters between two consecutive snapshots is fairly limited. Therefore, one intuitive way to track the MPCs/clusters is based on the distance, i.e., the MPCs pairs between two snapshots are associated if they have the smallest distance. Following this idea, a distance-based MPC tracking algorithm is proposed in \cite{Gan2015ClusterLC} to analyze the lifetime of the MPCs in V2V channels, where the MCD \cite{Steinbauer2002HowTQ} is used to measure the distance between the MPCs. Similarly, the MCD-based tracking method is also applied in \cite{Hanpinitsak2019FrequencyCO} and \cite{Materum2009MobileSS}. Instead of simply tracking the MPCs, \cite{Wang2017AFO} tracks the MPCs in the current snapshot to the cluster centroid in the last snapshot based on the MCD between the MPCs and the cluster centroids, as an initialization for the clustering algorithm. The algorithms above track the MPCs only based on the similarities without considering the physical environment, whereas \cite{Hanpinitsak2017MultipathCA} reconstructs the physical propagation environment and locates the reflection object based on the AoA and AoD, and then tracks the reflection point along the surface of the reflecting object. Fig. \ref{fig_tracking_dis} gives the key idea of this reflection point tracking. Apparently, the reflection point tracking is able to utilize the physical environment feature to improve the tracking performance, but it is mainly applicable for single-bounce reflections as it is difficult to acquire the specific propagation process of a multi-bounce reflection.

Due to the tracking principle, the distance-based method usually first links (tracks) the MPCs pairs with the smallest distance, which is a locally optimum solution rather than a globally optimum solution.

\subsubsection{Matching-Based Tracking}
A globally optimal solution to the tracking problem is difficult to obtain due to its non-convex nature. On the other hand, tracking MPCs between two snapshots is equivalent to a matching problem between two data sets. Inspired by this, a globally-optimum tracking solution is proposed in \cite{Huang2017ANT, Huang2017ANP}, where the tracking of MPCs is converted to a bipartite graph of a general matching problem: every node (i.e., MPC) in two subsets links to each other (i.e., the evolution between each MPC pair) and every link has its own weight. The matching problem is solved by using the Kuhn-Munkres algorithm \cite{Munkres1957AlgorithmsFT}. Similar solutions are also adopted in \cite{Huang2019APB, Huang2020ClusteringPE, Wang2018RandomCN}. Besides the Kuhn-Munkres algorithm, the Hungarian
algorithm \cite{Bazaraa1977LinearPA} has been also widely used for the matching problem, where the main difference is that the Hungarian algorithm sets all links as the same weight during the assignment. Based on the Hungarian algorithm, a cluster tracking method for millimeter-wave channels is proposed in \cite{Lai2019MethodologyFM}.

Compared to distance-based tracking, matching-based tracking can provide a tracking result with higher accuracy and better robustness according to the comparison in \cite{Huang2017ANT}. Nonetheless, due to the matching procedure, the matching-based tracking does not consider the splitting and merging behavior of the dynamic clusters and the birth/death of MPCs.

\subsubsection{Threshold-Based Tracking}
For time-varying channels, the clusters may show complicated evolving behavior, e.g., birth/death, moving, splitting, or merging. Neither the distance-based nor the matching-based tracking methods can well identify the splitting and merging behavior. Therefore, instead of linking MPCs to one another, a threshold-based tracking method is developed to link the MPCs one-to-many or many-to-one. Follow this idea, an automatic tracking algorithm is proposed in \cite{Czink2006ANA} to track the cluster centroid, where a preset threshold is built to identify the splitting and merging behavior of the cluster. A similar tracking method has been adopted in \cite{Karedal2009AGS, He2015ADW, Mahler2017TrackingOW} to track the MPCs in V2V channels.

Since the threshold-based method tracks the MPCs/clusters only based on a preset threshold, it has the lowest computation complexity. Yet, the threshold settings significantly impact the tracking results, and different measurement data may require different threshold settings.

\subsubsection{Prediction-Based Tracking}
The tracking methods introduced above inherently only consider the similarity of the MPCs/clusters between two snapshots, which means that the long-term evolving behavior of the MPCs/clusters is ignored. For the time-varying propagation channels, e.g., V2V channels, the MPCs/clusters may correspond to a specific interaction object, the long-term evolving behavior of the MPCs/clusters thus can contribute to the tracking of them. Inspired by this, some studies analyze the evolutionary history of the MPCs/clusters and predict the moving position (i.e., the MPC in the next snapshot) to ultimately improve the tracking accuracy. The Kalman filter \cite{Brown1983IntroductionTR} has been widely used for tracking and prediction problems, and uses a linear system function to learn the dynamic change of input data and predict the system in the future. A cluster centroid tracking algorithm based on the Kalman filter is firstly proposed in \cite{Czink2007TrackingTC} for the MIMO channels, where the cluster centroid is predicted and used as the initial position for the KPM algorithm. Similar solutions are also adopted in \cite{Gedschold2018TrackingBM, Wu2020ANP}. The Kalman filter is designed to predict one MPC/cluster; alternatively, the particle filter \cite{Arulampalam2002ATO} is developed to track multiple targets at the same time. Inspired by this, a particle filer-based tracking algorithm is proposed in \cite{Yin2008TrackingOT} to track the MPCs in time-varying channels.

The prediction-based tracking takes into account the long-term evolving behavior of the MPCs/cluster, yet it can only predict the potential position of the MPCs/cluster in the future. In this case, the prediction-based tracking can be applied with matching-based tracking, i.e., use a prediction method to learn MPCs evolution pattern and predict the possible moving position and then use a matching-based tracking method to associate the predicted position and the MPCs in the next snapshot, to achieve better tracking performance.

It can be found that with the development of cluster identification, the clustering procedure, and the tracking procedure tend to be jointly considered to better identify dynamic clusters in time-varying channels, i.e., evolution-based cluster identification. Based on the existing research, the evolution-based clustering can achieve better robustness and accuracy performance when processing the time-varying channels data, comparing with either tracking first and clustering afterward or clustering first and tracking afterward. Hence, it is a promising direction for the future development of cluster identification of time-varying channels. Besides, most of the existing clustering algorithms have no off-line training process, and the computational complexity thus is a critical point to implementations, which needs to be carefully considered for designing clustering method.
\section{AI/ML-Assisted Optimization and Modeling for Antenna Design to Improve Radio Propagation}\label{sec_antenna}

\subsection{ML-Based Transmit Antenna Selection}	
The simplest method to use channel information to adapt antenna properties for improved antenna-channel interaction is to rely on an antenna array to provide different spatial properties of the antenna system. Specifically, one can select between a finite set of spatial filters, made possible by either turning on/off different antenna elements (in the case of transmit antenna selection (TAS) \cite{heath2001antenna, molisch2004mimo}), fixed beams \cite{Butler1961BeamformingMS}, or antenna tilt, selecting different preformed beams covering the horizontal plane, or steering the beam in elevation \cite{wilson1992electrical}. In a generic sense, this approach is about connecting the antenna array outputs to the transmitter or receiver through a beamforming network, where the array weights can be set according to the desired selection functionality.

Among these applications, TAS has received the most attention in the literature \cite{joung2016machine, he2018transmit, cai2019antenna, zhang2020efficient, chen2019intelligent}, due to the topic being of interest since the early days of multi-antenna systems \cite{molisch2004mimo}. The basic idea of TAS is to devise an algorithm to select $P$ out of $M_{\rm T}$ transmitting antennas ($P < M_{\rm T}$) to be connected to the RF chains, such that the best possible end-to-end link performance is achieved. This technique is motivated both by the reduced hardware cost due to the need for fewer RF chains and the modest loss of performance under some conditions.

Optimal TAS algorithms can be computationally expensive, and their performance is dependent on the quality and completeness of the channel knowledge. An early paper that explores the use of pattern-recognition-based ML in TAS is \cite{joung2016machine}, where multiclass classification is attempted by building and applying a classification model from sufficient training data of the channel. Results obtained from simulated channels (2000 training samples, with correlations in both transmitting and receiving antennas) using multiclass-KNN and SVM algorithms are compared to conventional TAS using optimization-driven selection (maximize minimum eigenvalue or norm of the channel matrix \textbf{H} over all possible subsets). For single-stream transmission, all methods are equivalent in bit error rate (BER) performance. However, for two-stream transmission, the two ML methods cannot match the max-min eigenvalue method, but they benefit from lower computational complexity, i.e., polynomial as opposed to exhaustive combinatorial search. Furthermore, depending on the feature vector used in ML methods, the amount of feedback for TAS can be reduced relative to conventional methods with full-channel feedback. To further improve the performance, a new feature of the channel matrix is proposed in \cite{joung2016machine}, based on the norm square of the elements of $\textbf{H}^H\textbf{H}$ instead of \textbf{H}. The simulated results show that the proposed feature gives significantly better BER results for SVM in \cite{joung2016machine}.	

Following the same general idea in \cite{joung2016machine}, ML is investigated for a TAS setup in \cite{he2018transmit}, but for a different purpose, namely enhancing the secrecy rate of MIMO multi-antenna eavesdropper wiretap channels. In this case, Naive Bayesian (NB) is applied instead of KNN or SVM to provide a better secrecy rate with low complexity. The performance of SVM and NB slightly differ depending on the knowledge of the eavesdropper channel, but both can achieve nearly the same performance as the conventional method of exhaustive search. Moreover, SVM has lower complexity and both ML methods require half the amount of feedback relative to the conventional method.	To achieve superior link capacity to the data-driven ML approaches of \cite{joung2016machine, he2018transmit}, DL is proposed for TAS in \cite{cai2019antenna}. The proposed framework, based on CNN and the criterion of channel capacity, allows rich features to be extracted from the channel matrices and used to train powerful classifiers for selecting antennas. It is demonstrated that CNN outperforms SVM and KNN in terms of capacity loss and its variance, but at the cost of much higher computational complexity inherent to DL.

A general TAS framework is given in \cite{zhang2020efficient} for both neural networks and Gradient Boosting Decision Trees (GBDT). For GBDT, it also analyzes the relative importance of different features pattern from the norm square of the diagonal and off-diagonal elements of $\textbf{H}^H\textbf{H}$, concluding from empirical data that the diagonal values are twice as important for the classification task as the off-diagonal values. The comparisons in BER performance and complexity reveal that ANN and GBDT can both approach near-optimal performance, but the latter demonstrates a better BER performance-efficiency trade-off.

The interest in applying ML to TAS is not limited to single-user MIMO, but it has also been considered for multi-user massive MIMO systems \cite{chen2019intelligent}. In \cite{chen2019intelligent}, the self-supervised learning-based Monte Carlo Tree Search (MCTS) method is proposed to solve the antenna selection problem, using channel capacity as the key performance indicator. The components in the TAS system model are mapped to the basic elements of MCTS (action, tree state, and reward). Linear regression is used to obtain channel features and provide prediction to MCTS, facilitating the self-supervised learning process. Simulation results show high search efficiency with near-optimal performance, with the BER performance giving 1-dB gain over the greedy search selection method. The proposed method also achieves similar near-optimal BER performance as the search-based branch and bound (BAB) method \cite{mesleh2008spatial}, but with 50\% complexity.
\subsection{ML-Based Beam Selection and Antenna Tilt Optimization}
Apart from TAS, ML has also been applied to \emph{beam selection}, which is an important enabler of 5G millimeter wave technology, in providing the beamforming gain needed to compensate for the higher propagation losses at higher frequencies. 
The use of DL architectures is considered in \cite{klautau20185g} to perform millimeter-wave beam management for a vehicle-to-infrastructure (V2I) 5G environment. It was shown that the beam selection accuracy of the deep reinforcement learning algorithm approaches that of the optimum time slot (and beam) allocation obtained with dynamic programming. However, the focus of \cite{klautau20185g} is on the generation of realistic data sets involving channel evolution over time by combining a ray-tracing simulator and a traffic simulator. A different twist in the use of DL for millimeter-wave beam selection is explored in \cite{sim2020deep, 8734054, 8662770, 8503086, 8642397}, where sub-6 GHz channel information is used to reduce the exhaustively search-based beam sweeping overhead during initial access by as much as 80\%. The PDP as calculated from the channel is usually used as input to the DNN structure, which is based on the hypothesis that PDP can be considered as a fingerprint for UE position in a given environment.

Using ML to determine antenna tilt for coverage and interference optimization is also a popular subject in the literature \cite{guo2013spectral, dandanov2017dynamic, parera2018transferring}. 
In simple terms, a suitable antenna tilt can improve signal reception within a cell and reduce interference towards other cells, leading to a higher signal-to-interference-plus-noise ratio (SINR) received by the users and increased sum data rate in the network \cite{dandanov2017dynamic}. However, the traditional fixed-tilt strategy is not adequate for the complex coverage and interference problem in heterogeneous networks (HetNet) \cite{guo2013spectral}. In \cite{guo2013spectral},  a distributed reinforcement learning algorithm is proposed, which does not need a base station or network-wide knowledge of hotspot locations. In the simulation results in that paper, the Boltzmann exploration algorithm can achieve convergence to a near-optimal solution within limited iterations and improve the throughput fairness by 45-56\% and the energy efficiency by 21-47\%, as compared to fixed strategies.
It is shown in \cite{dandanov2017dynamic} that distributed reinforcement learning is also attractive for the antenna tilting for self-optimization of the RAN, even for homogeneous networks. Simulation results show a 30\% increase in the sum rate in an urban scenario when the tilt angle is optimized.

\subsection{Data-Driven Design of Antenna Patterns}
To benefit more from channel information than aiding in the selection of antenna/beam/tilt, the channel data can be more directly utilized in optimizing antenna-channel interaction to improve system performance. Since optimizing antenna-channel interaction is mainly about  ``far-field matching'', one can design the antenna pattern for some desired properties.

One such track is to adapt antenna properties using data-driven (or data-dependent) beamforming \cite{capon1969high}, which is interesting especially for the mm-wave-based applications in 6th generation (6G) communication system. 
In \cite{long2018data}, the CIR is processed by ML to synthesize the required array weights to optimize criteria such as minimum SINR, where hybrid beamforming is used in a millimeter-wave system to balance hardware cost (i.e., relating to the number of RF chains) and flexibility (in beamforming). Whereas the digital baseband part can utilize beamforming algorithms such as zero-forcing (ZF) to create arbitrary beam patterns, the flexibility in the analog part is limited to a set of phase-only beamforming weights. Nevertheless, the digital baseband component in hybrid beamforming gives the system more degrees of freedom to maximize sum rates as compared to a fixed beam-selection scheme \cite{klautau20185g}. However, since the classical approach to maximize sum rates with ZF beamforming involves several matrix inversions and hence is computationally intensive for real-time application, a low complexity near-optimal method is devised by posing the beamforming problem as a multiclass classification problem. Given adequate training data, SVM is used to obtain a statistical classification model, which is then used to select the optimal beam to maximize the sum rate. In a more direct approach of optimizing the far-field pattern of an antenna array to improve antenna-channel interaction, \cite{lecci2020machine} proposes the use of ML to optimize the locations of thinned (or sparse) array elements for network-level metrics (such as SINR) given a modest set of simulated channel data from the 3GPP Urban Microcell (UMi) scenario.

\begin{figure}[!t]
    \centering
    \includegraphics[width=0.43 \textwidth]{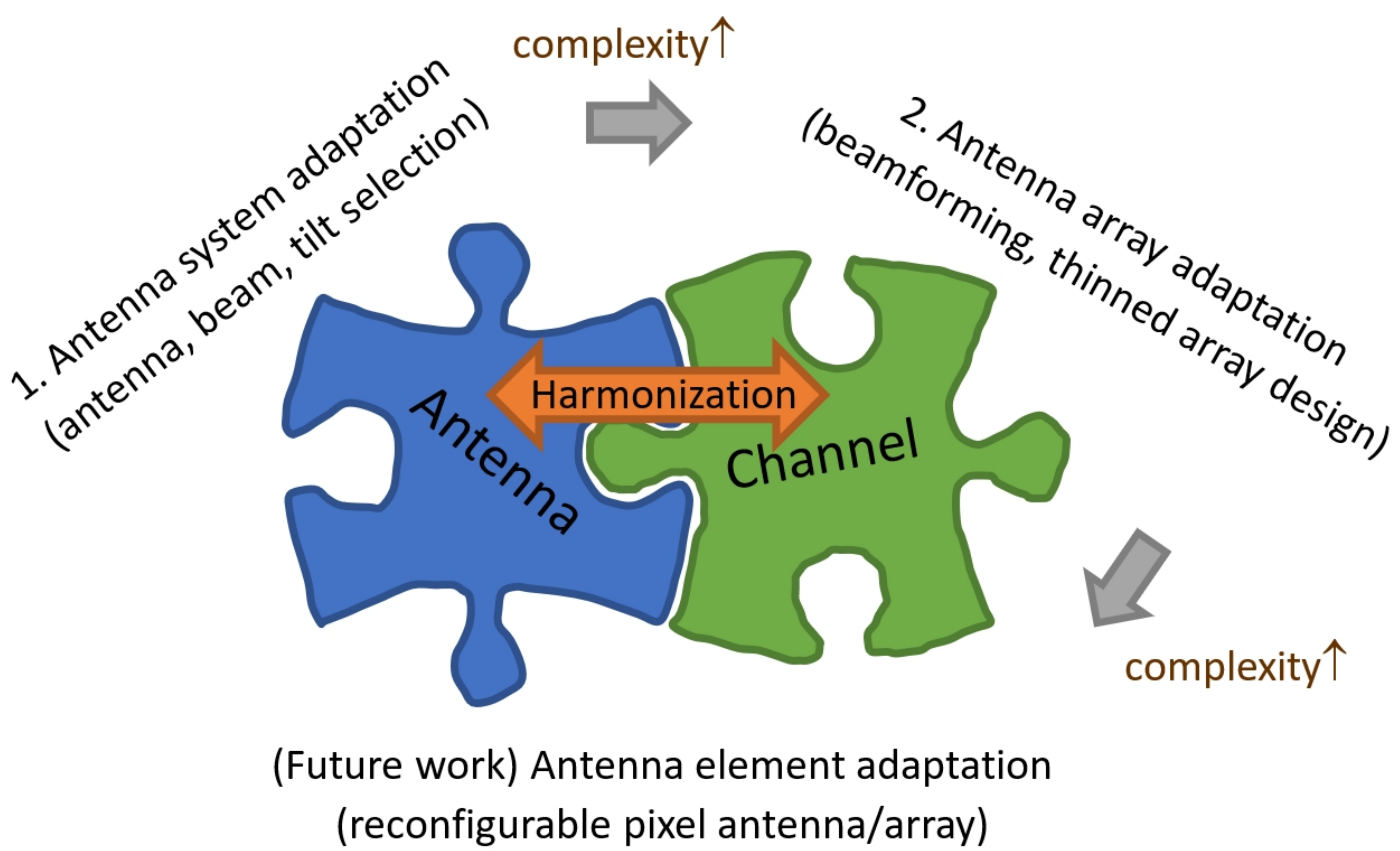}
    \caption{Illustration of optimization of antenna-channel interaction and different approaches to achieve it.}
    \label{fig_antenna_Channel}
    \vspace{-0.5 cm}
    \end{figure}

The downside of this more data-dependent approach is that it does not make use of the channel information to reconfigure the antenna elements in a more fundamental manner than adjusting antenna weights, and this limits the freedom for the antenna pattern to be synthesized for optimal antenna-channel interaction. To this end of optimizing antenna-channel interaction, two approaches have been attempted, differentiated by the level of sophistication, as illustrated in Fig. \ref{fig_antenna_Channel}:
the main research tracks for AI-based techniques for optimizing the compound of antennas and channels consist of either simple adaptation of some high-level antenna properties, such as antenna/beam/tilt selection, or more direct manipulation of antenna patterns, according to the channel data. It is foreseen that AI can be particularly useful for such an optimal pattern synthesis approach, as the future work given in Fig. \ref{fig_antenna_Channel}, since a flexible configuration of the radiating structure implies a large optimization search space.

\section{Important Issues and Challenges}
The challenges discussed in this section can be considered as guidelines for future ML-based channel analysis and modeling research.

\subsection{DL-Based Channel Parameter Estimation}
The channel feature extraction is a fundamental basis for channel analysis, which mainly consists of the channel parameter estimation and the cluster extraction. For most of the existing work, there are mainly two solutions to obtain the channel parameters: i) high-resolution parameter estimation method, e.g., SAGE or RiMAX; or ii)  Fourier transform-based filter, e.g., Bartlett Beamformer or Capon's Beamformer. The former methods can provide relatively more accurate parameters with the cost of high complexity caused by iterative computation, whereas the latter methods usually only provide a spectrum with relatively low computation complexity. Alternatively, the AI methods, especially ANN-based DL, have the potential to build the bridge between spectrum- and parameter-based solutions discussed in Section II.A: use the spectrum results as prior information and obtain the accurate channel parameters with fewer iterations. The main challenge here is i) how to efficiently utilize the spectrum result to reduce the training complexity; ii) how to improve the parameter estimation accuracy compared to the existing high-resolution parameter estimation algorithms but maintain the tradeoff between the training performance and overfitting.

\subsection{DL-Based Cluster Identification} For cluster recognition/extraction, most of the existing works rely on unsupervised clustering algorithms, e.g., K-Means or Fuzzy-C-Means. However, the unsupervised algorithms generally rely on pre-set parameters, e.g., the number and the position of initial cluster-centroids. Thus, the current clustering algorithm requires different pre-settings for different channel data, which requires extensive manual adjustment to maintain the clustering accuracy for non-stationary channels. On the other hand, the ANN-based DL shows great flexibility for the applications of target recognition and has already been extended to solve the clustering problem \cite{guo2017deep, ghasedi2017deep, huang2014deep}, which is highly related to the MPC's cluster recognition. Nevertheless, the accuracy of the DL-based cluster recognition is not increased as expected compared to the existing unsupervised clustering methods. Therefore, it requires more studies on how to further improve the accuracy and efficiency of the DL-based clustering methods. At the same time, the possibility of tracking joint clustering of time-varying MPCs also requires further investigation.

\subsection{AI-Based Antenna Design for Coverage Optimization or Energy Saving}
As reviewed in Section V, the main research tracks for AI-based antenna-channel optimization consist of either simple adaptation of some high-level antenna properties, such as antenna/beam/tilt selection, or more direct manipulation of antenna patterns, according to the channel data. Little effort has been made towards using AI techniques to enable the antenna pattern to be synthesized for optimal antenna-channel interaction by configuring the basic radiating structure in real-time. It is foreseen that AI can be particularly useful for such an optimal pattern synthesis approach since a flexible configuration of the radiating structure implies a large optimization search space. Furthermore, AI modeling can be used to predict the antenna pattern without having to run full-wave simulations for all possible configuration states. However, the existing basic building blocks for this future \cite{glazunov2009spherical,evans2010near} do not provide the means to translate a given pattern to the required antenna structure, nor do they give any indication of a reconfigurable antenna structure suitable for this purpose. AI-based modeling could be used to fill the gap between the patterns from full-wave simulations performed for a representative number of states and all possible patterns. Nevertheless, such a procedure is expected to be computationally very expensive. Then, multiple feeding ports will need to be positioned around the patch to excite the desired modes in the right proportion to generate a given optimal pattern. Therefore, this exercise, if performed in a brute-force manner, is still too complex for real-time application, which opens up the opportunity for AI-based modeling and utilization. Furthermore, appropriate impedance matching of the designed feeding structure will be needed, and this aspect will require further investigations.

\subsection{AI-Based Applications for mmWave/THz Band Communications}
Channel measurements (especially in mmWave and THz band) are often accompanied by maddening time consumption, engineering problems, and capital costs. The ML method needs training data, and reliable training data should come from the measurements in actual scenarios. In this sense, data acquisition is the bottleneck of many ML-based applications. Admittedly, some simulation methods can generate synthetic training data, but the simulation methods themselves also need measurement data for evaluation and verification. Hence, conducting sufficient measurement campaigns to support AI-based applications is one challenging aspect in the future for mmWave/THz channels. Meanwhile, the mmWave/THz channels with the ultra-wideband and ultra-massive MIMO have shown some new properties, e.g., channel sparsity, channel hardening, non-stationarity in time/spatial/frequency domains. These new channel properties may significantly affect channel data processing and have not been considered yet in the existing AI-based applications, e.g., channel sparsity property may contribute to cluster identification; channel hardening may improve the scenario identification. How to exploit new channel properties to improve the efficiency, accuracy, and robustness of AI-based applications in communications still requires further investigation.

\section{Conclusion}
AI techniques have become a necessary tool to develop the next generation communication network. In this paper, we provide a comprehensive overview of AI-enabled data processing for propagation channel studies, including channel parameter estimation and characterization and antenna-channel optimization in Part I, whereas the scenario identification and channel modeling/prediction are covered in Part II \cite{PartII}. This paper demonstrates the early results of the related works and illustrates the typical AI/ML-based solutions for each topic. Based on the state-of-art, the future challenges of AI/ML-based channel data processing techniques are given as well.

\bibliographystyle{ieeetr}
\bibliography{Overview_ref}

\begin{thebibliography}{100}

\bibitem{wang2020artificial}
C.-X. Wang, M.~Di~Renzo, S.~Stanczak, S.~Wang, and E.~G. Larsson,
  ``{Artificial} intelligence enabled wireless networking for {5G} and beyond:
  Recent advances and future challenges,'' {\em IEEE Wireless Commun.},
  vol.~27, no.~1, pp.~16--23, 2020.

\bibitem{tataria20216g}
H.~Tataria, M.~Shafi, A.~F. Molisch, M.~Dohler, H.~Sj{\"o}land, and
  F.~Tufvesson, ``{6G} wireless systems: Vision, requirements, challenges,
  insights, and opportunities,'' {\em Proc. IEEE}, 2021.

\bibitem{wang20206g}
C.-X. Wang, J.~Huang, H.~Wang, X.~Gao, X.~You, and Y.~Hao, ``{6G} wireless
  channel measurements and models: Trends and challenges,'' {\em IEEE Veh.
  Technol. Mag.}, vol.~15, no.~4, pp.~22--32, 2020.

\bibitem{7999256}
T.~S. Rappaport, G.~R. MacCartney, S.~Sun, H.~Yan, and S.~Deng, ``Small-scale,
  local area, and transitional millimeter wave propagation for {5G}
  communications,'' {\em IEEE Trans. Antennas Propag.}, vol.~65, no.~12,
  pp.~6474--6490, 2017.

\bibitem{khuwaja2018survey}
A.~A. Khuwaja, Y.~Chen, N.~Zhao, M.-S. Alouini, and P.~Dobbins, ``A survey of
  channel modeling for {UAV} communications,'' {\em IEEE Commun. Surv. \&
  Tutorials}, vol.~20, no.~4, pp.~2804--2821, 2018.

\bibitem{fan2018wireless}
W.~Fan, F.~Zhang, P.~Ky{\"o}sti, L.~Hentil{\"a}, and G.~F. Pedersen, ``Wireless
  cable method for high-order {MIMO} terminals based on particle swarm
  optimization algorithm,'' {\em IEEE Trans. Antennas Propag.}, vol.~66,
  no.~10, pp.~5536--5545, 2018.

\bibitem{9362269}
Y.~Tan, C.-X. Wang, J.~{\O}. Nielsen, G.~F. Pedersen, and Q.~Zhu, ``A novel
  {B5G} frequency non-stationary wireless channel model,'' {\em IEEE Trans.
  Antennas Propag.}, pp.~1--1, 2021.

\bibitem{686758}
S.-S. Jeng, G.~Okamoto, G.~Xu, H.-P. Lin, and W.~Vogel, ``Experimental
  evaluation of smart antenna system performance for wireless communications,''
  {\em IEEE Trans. Antennas Propag.}, vol.~46, no.~6, pp.~749--757, 1998.

\bibitem{4052596}
A.~Alomainy, Y.~Hao, A.~Owadally, C.~G. Parini, Y.~Nechayev, C.~C.
  Constantinou, and P.~S. Hall, ``Statistical analysis and performance
  evaluation for on-body radio propagation with microstrip patch antennas,''
  {\em IEEE Trans. Antennas Propag.}, vol.~55, no.~1, pp.~245--248, 2007.

\bibitem{6280628}
H.~P. Bui, Y.~Ogawa, T.~Nishimura, and T.~Ohgane, ``Performance evaluation of a
  multi-user {MIMO} system with prediction of time-varying indoor channels,''
  {\em IEEE Trans. Antennas Propag.}, vol.~61, no.~1, pp.~371--379, 2013.

\bibitem{575640}
J.~Rossi, J.-P. Barbot, and A.~Levy, ``Theory and measurement of the angle of
  arrival and time delay of uhf radiowaves using a ring array,'' {\em IEEE
  Trans. Antennas Propag.}, vol.~45, no.~5, pp.~876--884, 1997.

\bibitem{bi2015wireless}
S.~Bi, R.~Zhang, Z.~Ding, and S.~Cui, ``Wireless communications in the era of
  big data,'' {\em IEEE Commun. Mag.}, vol.~53, no.~10, pp.~190--199, 2015.

\bibitem{chen2019artificial}
M.~Chen, U.~Challita, W.~Saad, C.~Yin, and M.~Debbah, ``Artificial neural
  networks-based machine learning for wireless networks: A tutorial,'' {\em
  IEEE Commun. Surv. \& Tutorials}, vol.~21, no.~4, pp.~3039--3071, 2019.

\bibitem{bin2020machine}
H.~Bin and H.~D. SCHOTTEN, ``Machine learning for network slicing resource
  management: A comprehensive survey,'' {\em ZTE Communications}, vol.~17,
  no.~4, pp.~27--32, 2020.

\bibitem{songyan2020learn}
X.~Songyan, L.~Ang, W.~Jinfei, Y.~Na, M.~Yi, R.~TAFAZOLLI, and T.~DODGSON, ``To
  learn or not to learn: Deep learning assisted wireless modem design,'' {\em
  ZTE Communications}, vol.~17, no.~4, pp.~3--11, 2020.

\bibitem{lin2020adaptive}
S.~Lin, D.~Jiangbing, H.~Feng, T.~Ningfeng, and H.~Zuyuan, ``Adaptive and
  intelligent digital signal processing for improved optical interconnection,''
  {\em ZTE Communications}, vol.~18, no.~2, pp.~57--73, 2020.

\bibitem{MBOOK}
R.~He and Z.~Ding, {\em Applications of Machine Learning in Wireless
  Communications}.
\newblock Institution of Engineering and Technol., July 2019.

\bibitem{8952905}
Q.~Wu, H.~Wang, and W.~Hong, ``Multistage collaborative machine learning and
  its application to antenna modeling and optimization,'' {\em IEEE Trans.
  Antennas Propag.}, vol.~68, no.~5, pp.~3397--3409, 2020.

\bibitem{haenlein2019brief}
M.~Haenlein and A.~Kaplan, ``A brief history of artificial intelligence: On the
  past, present, and future of artificial intelligence,'' {\em California
  management review}, vol.~61, no.~4, pp.~5--14, 2019.

\bibitem{dietterich1997machine}
T.~G. Dietterich, ``Machine-learning research,'' {\em AI magazine}, vol.~18,
  no.~4, pp.~97--97, 1997.

\bibitem{goodfellow2016deep}
I.~Goodfellow, Y.~Bengio, A.~Courville, and Y.~Bengio, {\em Deep learning},
  vol.~1.
\newblock MIT press Cambridge, 2016.

\bibitem{shrestha2019review}
A.~Shrestha and A.~Mahmood, ``Review of deep learning algorithms and
  architectures,'' {\em IEEE Access}, vol.~7, pp.~53040--53065, 2019.

\bibitem{he2021wireless}
R.~He, B.~Ai, G.~Wang, M.~Yang, and Z.~Zhong, ``Wireless channel sparsity:
  Measurement, analysis, and exploitation in estimation,'' {\em IEEE Wireless
  Commun.}, vol.~PP, no.~99, pp.~1--7, 2021.

\bibitem{9395374}
Q.~Wu, W.~Chen, C.~Yu, H.~Wang, and W.~Hong, ``Multilayer machine
  learning-assisted optimization-based robust design and its applications to
  antennas and arrays,'' {\em IEEE Trans. Antennas Propag.}, pp.~1--1, 2021.

\bibitem{9158524}
J.~Zhang, M.~O. Akinsolu, B.~Liu, and G.~A.~E. Vandenbosch, ``Automatic
  {AI}-driven design of mutual coupling reducing topologies for frequency
  reconfigurable antenna arrays,'' {\em IEEE Trans. Antennas Propag.}, vol.~69,
  no.~3, pp.~1831--1836, 2021.

\bibitem{9212600}
D.~Xiao, L.~Guo, W.~Liu, and M.~Hou, ``Efficient {RCS} prediction of the
  conducting target based on physics-inspired machine learning and experimental
  design,'' {\em IEEE Trans. Antennas Propag.}, vol.~69, no.~4, pp.~2274--2289,
  2021.

\bibitem{PartII}
C.~Huang, R.~He, B.~Ai, A.~F. Molisch, B.~K. Lau, K.~Haneda, B.~Liu, C.-X.
  Wang, M.~Yang, O.~Claude, and Z.~Zhong, ``Artificial intelligence enabled
  radio propagation for communications---{Part II}: Scenario identification and
  channel modeling,'' {\em IEEE Trans. Antennas Propag.}, submitted.

\bibitem{6060881}
M.~Landmann, M.~Kaske, and R.~S. Thoma, ``Impact of incomplete and inaccurate
  data models on high resolution parameter estimation in multidimensional
  channel sounding,'' {\em IEEE Trans. Antennas Propag.}, vol.~60, no.~2,
  pp.~557--573, 2012.

\bibitem{9104014}
X.~Cai, W.~Fan, X.~Yin, and G.~F. Pedersen, ``Trajectory-aided
  maximum-likelihood algorithm for channel parameter estimation in
  ultrawideband large-scale arrays,'' {\em IEEE Trans. Antennas Propag.},
  vol.~68, no.~10, pp.~7131--7143, 2020.

\bibitem{5979196}
J.~Salmi and A.~F. Molisch, ``Propagation parameter estimation, modeling and
  measurements for ultrawideband {MIMO} radar,'' {\em IEEE Trans. Antennas
  Propag.}, vol.~59, no.~11, pp.~4257--4267, 2011.

\bibitem{299557}
E.~Dowling, R.~DeGroat, and D.~Linebarger, ``Exponential parameter estimation
  in the presence of known components and noise,'' {\em IEEE Trans. Antennas
  Propag.}, vol.~42, no.~5, pp.~590--599, 1994.

\bibitem{8852816}
M.~Ershadh and M.~Meenakshi, ``A new modeling methodology for multipath
  parameter estimation in ultrawideband channels,'' {\em IEEE Trans. Antennas
  Propag.}, vol.~69, no.~4, pp.~2249--2255, 2021.

\bibitem{turin1972statistical}
G.~L. Turin, F.~D. Clapp, T.~L. Johnston, S.~B. Fine, and D.~Lavry, ``A
  statistical model of urban multipath propagation,'' {\em IEEE Trans. Veh.
  Technol.}, vol.~21, no.~1, pp.~1--9, 1972.

\bibitem{Saleh1986ASM}
A.~Saleh and R.~Valenzuela, ``A statistical model for indoor multipath
  propagation,'' {\em IEEE J. Sel. Areas Commun.}, vol.~5, pp.~128--137, 1986.

\bibitem{spencer2000modeling}
Q.~H. Spencer, B.~D. Jeffs, M.~A. Jensen, and A.~L. Swindlehurst, ``Modeling
  the statistical time and angle of arrival characteristics of an indoor
  multipath channel,'' {\em IEEE J. Sel. Areas Commun.}, vol.~18, no.~3,
  pp.~347--360, 2000.

\bibitem{Fuhl1998UnifiedCM}
J.~Fuhl, A.~Molisch, and E.~Bonek, ``Unified channel model for mobile radio
  systems with smart antennas,'' in {\em Proc. IEE P-RADAR SON NAV'1998},
  pp.~32--41, IEE, 1998.

\bibitem{Molisch2006TheCD}
A.~Molisch, H.~Asplund, R.~Heddergott, M.~Steinbauer, and T.~Zwick, ``The
  {COST} 259 directional channel model-{Part I}: Overview and methodology,''
  {\em IEEE Trans. Wireless Commun.}, vol.~5, 2006.

\bibitem{Asplund2006TheC2}
H.~Asplund, A.~Glazunov, A.~Molisch, K.~Pedersen, and M.~Steinbauer, ``The
  {COST} 259 directional channel model-{Part II}: Macrocells,'' {\em IEEE
  Trans. Wireless Commun.}, vol.~5, 2006.

\bibitem{Liu2012TheC2}
L.~Liu, C.~Oestges, J.~Poutanen, K.~Haneda, P.~Vainikainen, F.~Quitin,
  F.~Tufvesson, and P.~Doncker, ``The {COST} 2100 {MIMO} channel model,'' {\em
  IEEE Wireless Commun.}, vol.~19, 2012.

\bibitem{3gpp}
{3GPP TR 25.996}, {\em Spatial channel model for multiple input multiple output
  (MIMO) simulations}, Jul. 2012.

\bibitem{Meinila2009WINNERIC}
C.~Schneider, M.~Narandzic, M.~K{\"a}ske, G.~Sommerkorn, and R.~S. Thom{\"a},
  ``Large scale parameter for the {WINNER} {II} channel model at 2.53 {GHz} in
  urban macro cell,'' in {\em Proc. IEEE VTC'2010}, pp.~1--5, IEEE, 2010.

\bibitem{KystiWINNERIC}
Y.~d.~J. Bultitude and T.~Rautiainen, ``Ist-4-027756 winner ii d1. 1.2 v1. 2
  winner ii channel models,'' {\em EBITG, TUI, UOULU, CU/CRC, NOKIA, Tech.
  Rep}, 2007.

\bibitem{Karedal2007AMS}
J.~Karedal, S.~Wyne, P.~Almers, F.~Tufvesson, and A.~Molisch, ``A
  measurement-based statistical model for industrial ultra-wideband channels,''
  {\em IEEE Trans. Wireless Commun.}, vol.~6, 2007.

\bibitem{Cramer2002EvaluationOA}
R.~Cramer, R.~Scholtz, and M.~Win, ``Evaluation of an ultra-wide-band
  propagation channel,'' {\em IEEE Trans. Antennas Propag.}, vol.~50,
  pp.~561--570, 2002.

\bibitem{Czink2007ClusterCI}
N.~Czink, X.~Yin, H.~Ozcelik, M.~Herdin, E.~Bonek, and B.~Fleury, ``Cluster
  characteristics in a {MIMO} indoor propagation environment,'' {\em IEEE
  Trans. Wireless Commun.}, vol.~6, 2007.

\bibitem{Chong2005AGS}
C.-C. Chong and S.~K. Yong, ``A generic statistical-based {UWB} channel model
  for high-rise apartments,'' {\em IEEE Trans. Antennas Propag.}, vol.~53,
  pp.~2389--2399, 2005.

\bibitem{Chen2019ArtificialNN}
M.~Chen, U.~Challita, W.~Saad, C.~Yin, and M.~Debbah, ``Artificial neural
  networks-based machine learning for wireless networks: A tutorial,'' {\em
  IEEE Commun. Surv. \& Tutorials}, vol.~21, pp.~3039--3071, 2019.

\bibitem{Huang2019MachineLearningBasedDP}
C.~Huang, A.~Molisch, R.~He, R.~Wang, P.~Tang, and Z.~Zhong,
  ``Machine-learning-based data processing techniques for vehicle-to-vehicle
  channel modeling,'' {\em IEEE Commun. Mag.}, vol.~57, pp.~109--115, 2019.

\bibitem{Jain1999DataCA}
A.~K. Jain, M.~Murty, and P.~Flynn, ``Data clustering: a review,'' {\em ACM
  Comput. Surv.}, vol.~31, pp.~264--323, 1999.

\bibitem{he2018clustering}
R.~He, B.~Ai, A.~F. Molisch, G.~L. Stuber, Q.~Li, Z.~Zhong, and J.~Yu,
  ``Clustering enabled wireless channel modeling using big data algorithms,''
  {\em IEEE Commun. Mag.}, vol.~56, no.~5, pp.~177--183, 2018.

\bibitem{fan2013emulating}
W.~Fan, F.~Sun, J.~{\O}. Nielsen, M.~B. Knudsen, G.~F. Pedersen, {\em et~al.},
  ``Emulating spatial characteristics of {MIMO} channels for {OTA} testing,''
  {\em IEEE Trans. Antennas Propag.}, vol.~61, no.~8, pp.~4306--4314, 2013.

\bibitem{fan2013antenna}
W.~Fan, J.~{\O}. Nielsen, O.~Franek, X.~Carre{\~n}o, J.~S. Ashta, M.~B.
  Knudsen, and G.~F. Pedersen, ``Antenna pattern impact on {MIMO} {OTA}
  testing,'' {\em IEEE Trans. Antennas Propag.}, vol.~61, no.~11,
  pp.~5714--5723, 2013.

\bibitem{6863647}
M.~{\'A}. Garc{\'\i}a-Fern{\'a}ndez, D.~Carsenat, and C.~Decroze, ``Antenna
  gain and radiation pattern measurements in reverberation chamber using
  doppler effect,'' {\em IEEE Trans. Antennas Propag.}, vol.~62, no.~10,
  pp.~5389--5394, 2014.

\bibitem{7875483}
S.~Zhang, X.~Chen, I.~Syrytsin, and G.~F. Pedersen, ``A planar switchable
  {3-D}-coverage phased array antenna and its user effects for 28-{GHz} mobile
  terminal applications,'' {\em IEEE Trans. Antennas Propag.}, vol.~65, no.~12,
  pp.~6413--6421, 2017.

\bibitem{8691774}
T.~Fukasawa, N.~Yoneda, and H.~Miyashita, ``Investigation on current reduction
  effects of baluns for measurement of a small antenna,'' {\em IEEE Trans.
  Antennas Propag.}, vol.~67, no.~7, pp.~4323--4329, 2019.

\bibitem{heath2001antenna}
R.~W. Heath, S.~Sandhu, and A.~Paulraj, ``Antenna selection for spatial
  multiplexing systems with linear receivers,'' {\em IEEE Commun. Lett.},
  vol.~5, no.~4, pp.~142--144, 2001.

\bibitem{molisch2004mimo}
A.~F. Molisch and M.~Z. Win, ``{MIMO} systems with antenna selection,'' {\em
  IEEE Microwave Mag.}, vol.~5, no.~1, pp.~46--56, 2004.

\bibitem{joung2016machine}
J.~Joung, ``Machine learning-based antenna selection in wireless
  communications,'' {\em IEEE Commun. Lett.}, vol.~20, no.~11, pp.~2241--2244,
  2016.

\bibitem{he2018transmit}
D.~He, C.~Liu, T.~Q. Quek, and H.~Wang, ``Transmit antenna selection in {MIMO}
  wiretap channels: A machine learning approach,'' {\em IEEE Wireless Commun.
  Letters}, vol.~7, no.~4, pp.~634--637, 2018.

\bibitem{cai2019antenna}
J.-x. Cai, R.~Zhong, and Y.~Li, ``Antenna selection for multiple-input
  multiple-output systems based on deep convolutional neural networks,'' {\em
  PloS one}, vol.~14, no.~5, p.~e0215672, 2019.

\bibitem{zhang2020efficient}
Y.~Zhang, J.~Wang, X.~Wang, Y.~Xue, and J.~Song, ``Efficient selection on
  spatial modulation antennas: Learning or boosting,'' {\em IEEE Wireless
  Commun. Letters}, vol.~9, no.~8, pp.~1249--1252, 2020.

\bibitem{chen2019intelligent}
J.~Chen, S.~Chen, Y.~Qi, and S.~Fu, ``Intelligent massive {MIMO} antenna
  selection using {Monte} {Carlo} tree search,'' {\em IEEE Trans. Sig. Proc.},
  vol.~67, no.~20, pp.~5380--5390, 2019.

\bibitem{bartlett1948smoothing}
M.~S. Bartlett, ``Smoothing periodograms from time-series with continuous
  spectra,'' {\em Nature}, vol.~161, no.~4096, pp.~686--687, 1948.

\bibitem{schmidt1986multiple}
R.~Schmidt, ``Multiple emitter location and signal parameter estimation,'' {\em
  IEEE Trans. Antennas Propag.}, vol.~34, no.~3, pp.~276--280, 1986.

\bibitem{roy1986estimation}
R.~Roy, A.~Paulraj, and T.~Kailath, ``Estimation of signal parameters via
  rotational invariance techniques-{ESPRIT},'' in {\em Proc. IEEE MILCOM'1986},
  vol.~3, pp.~41--6, IEEE, 1986.

\bibitem{fleury1999channel}
B.~H. Fleury, M.~Tschudin, R.~Heddergott, D.~Dahlhaus, and K.~I. Pedersen,
  ``Channel parameter estimation in mobile radio environments using the {SAGE}
  algorithm,'' {\em IEEE J. Sel. Areas Commun.}, vol.~17, no.~3, pp.~434--450,
  1999.

\bibitem{thoma2004multidimensional}
R.~S. Thoma, M.~Landmann, G.~Sommerkorn, and A.~Richter, ``Multidimensional
  high-resolution channel sounding in mobile radio,'' in {\em Proc. IEEE
  I2MTC'2004}, vol.~1, pp.~257--262, IEEE, 2004.

\bibitem{uccellari2017application}
M.~Uccellari, F.~Facchini, M.~Sola, E.~Sirignano, G.~M. Vitetta, A.~Barbieri,
  and S.~Tondelli, ``On the application of support vector machines to the
  prediction of propagation losses at 169 {MHz} for smart metering
  applications,'' {\em IET Microwaves Antennas \& Propag.}, vol.~12, no.~3,
  pp.~302--312, 2017.

\bibitem{bai2019prediction}
L.~Bai, C.-X. Wang, Q.~Xu, S.~Ventouras, and G.~Goussetis, ``Prediction of
  channel excess attenuation for satellite communication systems at {Q}-band
  using artificial neural network,'' {\em IEEE Antennas Wirel. \& Propag.
  Lett.}, vol.~18, no.~11, pp.~2235--2239, 2019.

\bibitem{yang2021machine}
M.~Yang, B.~Ai, R.~He, C.~Huang, Z.~Ma, Z.~Zhong, J.~Wang, L.~Pei, Y.~Li, and
  J.~Li, ``Machine-learning-based fast angle-of-arrival recognition for
  vehicular communications,'' {\em IEEE Trans. Veh. Technol.}, 2021.

\bibitem{liu2012efficient}
Z.-M. Liu, Z.-T. Huang, and Y.-Y. Zhou, ``An efficient maximum likelihood
  method for direction-of-arrival estimation via sparse bayesian learning,''
  {\em IEEE Trans. Wireless Commun.}, vol.~11, no.~10, pp.~1--11, 2012.

\bibitem{cal2013relevance}
J.~A. Cal-Braz, L.~J. Matos, and E.~Cataldo, ``The relevance vector machine
  applied to the modeling of wireless channels,'' {\em IEEE Trans. Antennas
  Propag.}, vol.~61, no.~12, pp.~6157--6167, 2013.

\bibitem{4732313}
J.~{Salmi}, A.~{Richter}, and V.~{Koivunen}, ``Detection and tracking of {MIMO}
  propagation path parameters using state-space approach,'' {\em IEEE Trans.
  Signal Process.}, vol.~57, no.~4, pp.~1538--1550, 2009.

\bibitem{huang2018big}
J.~Huang, C.-X. Wang, L.~Bai, J.~Sun, Y.~Yang, J.~Li, O.~Tirkkonen, and M.-T.
  Zhou, ``A big data enabled channel model for {5G} wireless communication
  systems,'' {\em IEEE Trans. Big Data}, vol.~6, no.~2, pp.~211--222, 2018.

\bibitem{huang2018deep}
H.~Huang, J.~Yang, H.~Huang, Y.~Song, and G.~Gui, ``Deep learning for
  super-resolution channel estimation and {DOA} estimation based massive {MIMO}
  system,'' {\em IEEE Trans. Veh. Technol.}, vol.~67, no.~9, pp.~8549--8560,
  2018.

\bibitem{fan2018flexible}
W.~Fan, P.~Ky{\"o}sti, L.~Hentil{\"a}, and G.~F. Pedersen, ``A flexible
  millimeter-wave radio channel emulator design with experimental
  validations,'' {\em IEEE Trans. Antennas Propag.}, vol.~66, no.~11,
  pp.~6446--6451, 2018.

\bibitem{Jaeckel14TAP}
S.~Jaeckel, L.~Raschkowski, K.~B{\"o}rner, and L.~Thiele, ``{QuaDRiGa}: A {3-D}
  multi-cell channel model with time evolution for enabling virtual field
  trials,'' {\em IEEE Trans. Antennas Propag.}, vol.~62, pp.~3242--3256, June
  2014.

\bibitem{Remcom}
R.~Eichenlaub, C.~Valentine, S.~Fast, and S.~Albarano, ``Fidelity at high
  speed: Wireless insite$\textregistered$ real time module$\texttrademark$,''
  in {\em Proc. IEEE MILCOM'2008}, pp.~1--7, 2008.

\bibitem{Shutin2004ClusterAO}
D.~Shutin and G.~Kubin, ``Cluster analysis of wireless channel impulse
  responses with hidden markov models,'' in {\em Proc. IEEE ICASS'2004},
  vol.~4, pp.~iv--949, IEEE, 2004.

\bibitem{Rabiner1989ATO}
L.~Rabiner, ``A tutorial on hidden markov models and selected applications,''
  {\em Proc. IEEE}, 1989.

\bibitem{Chuang2007AutomatedIO}
J.~Chuang, S.~Bashir, and D.~Michelson, ``Automated identification of clusters
  in {UWB} channel impulse responses,'' {\em 2007 Canadian Conference on
  Electrical and Computer Engineering}, pp.~761--764, 2007.

\bibitem{Corrigan2009AutomaticUC}
M.~Corrigan, A.~Walton, W.~Niu, J.~Li, and T.~Talty, ``Automatic {UWB} clusters
  identification,'' {\em 2009 IEEE Radio and Wireless Symposium}, pp.~376--379,
  2009.

\bibitem{Gentile2013UsingTK}
C.~Gentile, ``Using the kurtosis measure to identify clusters in wireless
  channel impulse responses,'' {\em IEEE Trans. Antennas Propag.}, vol.~61,
  pp.~3392--3395, 2013.

\bibitem{Zhu1996RegionCU}
S.~Zhu and A.~Yuille, ``Region competition: Unifying snakes, region growing,
  and bayes/{MDL} for multiband image segmentation,'' {\em IEEE Trans. Pattern
  Anal. Mach. Intell.}, vol.~18, pp.~884--900, 1996.

\bibitem{He2016OnTC}
R.~He, W.~Chen, B.~Ai, A.~Molisch, W.~Wang, Z.~Zhong, J.~Yu, and S.~Sangodoyin,
  ``On the clustering of radio channel impulse responses using sparsity-based
  methods,'' {\em IEEE Trans. Antennas Propag.}, vol.~64, pp.~2465--2474, 2016.

\bibitem{He2016ASC}
R.~He, W.~Chen, B.~Ai, A.~Molisch, W.~Wang, Z.~Zhong, J.~Yu, and S.~Sangodoyin,
  ``A sparsity-based clustering framework for radio channel impulse
  responses,'' {\em 2016 IEEE 83rd Vehicular Technology Conference (VTC
  Spring)}, pp.~1--5, 2016.

\bibitem{Li2020ClusteringAI}
Y.~Li, J.~Zhang, Z.~Ma, and Y.~Zhang, ``Clustering analysis in the wireless
  propagation channel with a variational gaussian mixture model,'' {\em IEEE
  Trans. Big Data}, vol.~6, pp.~223--232, 2020.

\bibitem{Li2018ClusteringIW}
Y.~Li, J.~Zhang, and Z.~Ma, ``Clustering in wireless propagation channel with a
  statistics-based framework,'' {\em 2018 IEEE Wireless Commun. and Networking
  Conference (WCNC)}, pp.~1--6, 2018.

\bibitem{Tzikas2008TheVA}
D.~G. Tzikas, A.~C. Likas, and N.~P. Galatsanos, ``The variational
  approximation for bayesian inference,'' {\em IEEE Signal Process Mag.},
  vol.~25, pp.~131--146, 2008.

\bibitem{Ma2011BayesianEO}
Z.~Ma and A.~Leijon, ``Bayesian estimation of beta mixture models with
  variational inference,'' {\em IEEE Trans. Pattern Anal. Mach. Intell.},
  vol.~33, pp.~2160--2173, 2011.

\bibitem{Salo2005AutomaticCO}
J.~Salo, J.~Salmi, N.~Czink, and P.~Vainikainen, ``Automatic clustering of
  nonstationary {MIMO} channel parameter estimates,'' in {\em Proc. na},
  pp.~1--5.

\bibitem{Steinbauer2002HowTQ}
M.~Steinbauer, H.~Ozcelik, H.~Hofstetter, C.~Mecklenbrauker, and E.~Bonek,
  ``How to quantify multipath separation,'' {\em IEICE Trans. Electron.},
  vol.~85, pp.~552--557, 2002.

\bibitem{Czink2006ImprovingCP}
N.~Czink, P.~Cera, J.~Salo, E.~Bonek, J.~Nuutinen, and J.~Ylitalo, ``Improving
  clustering performance using multipath component distance,'' {\em Electronics
  Letters}, vol.~42, pp.~33--35, 2006.

\bibitem{Czink2005AutomaticCO}
N.~Czink, P.~Cera, J.~Salo, E.~Bonek, J.~Nuutinen, and J.~Ylitalo, ``Automatic
  clustering of {MIMO} channel parameters using the multi-path component
  distance measure,'' in {\em Proc. Citeseer}, pp.~1--5, 2005.

\bibitem{Czink2006AFF}
N.~Czink, P.~Cera, J.~Salo, E.~Bonek, J.~Nuutinen, and J.~Ylitalo, ``A
  framework for automatic clustering of parametric {MIMO} channel data
  including path powers,'' {\em IEEE Vehicular Technology Conference},
  pp.~1--5, 2006.

\bibitem{1017616}
T.~{Kanungo}, D.~M. {Mount}, N.~S. {Netanyahu}, C.~D. {Piatko}, R.~{Silverman},
  and A.~Y. {Wu}, ``An efficient k-means clustering algorithm: analysis and
  implementation,'' {\em IEEE Trans. Pattern Anal. Mach. Intell.}, vol.~24,
  no.~7, pp.~881--892, 2002.

\bibitem{mota2011clustering}
S.~Mota, M.~O. Garcia, A.~Rocha, and F.~P{\'e}rez-Font{\'a}n, ``Clustering of
  the multipath radio channel parameters,'' in {\em Proc. EuCAP'2011},
  pp.~3232--3236, IEEE, 2011.

\bibitem{Schneider2009ClusteringOM}
C.~Schneider, M.~Bauer, M.~Narandzic, W.~Kotterman, and R.~Thom{\"a},
  ``Clustering of {MIMO} channel parameters - performance comparison,'' {\em
  VTC Spring 2009 - IEEE 69th Vehicular Technology Conference}, pp.~1--5, 2009.

\bibitem{Samimi20153DSC}
M.~K. Samimi and T.~S. Rappaport, ``{3-D} statistical channel model for
  millimeter-wave outdoor mobile broadband communications,'' in {\em Proc. IEEE
  ICC'2015}, pp.~2430--2436, IEEE, 2015.

\bibitem{Samimi20163DMS}
M.~Samimi and T.~S. Rappaport, ``{3-D} millimeter-wave statistical channel
  model for {5G} wireless system design,'' {\em IEEE Trans. Microwave Theory
  Tech.}, vol.~64, pp.~2207--2225, 2016.

\bibitem{Ester1996ADA}
M.~Ester, H.-P. Kriegel, J.~Sander, X.~Xu, {\em et~al.}, ``A density-based
  algorithm for discovering clusters in large spatial databases with noise.,''
  in {\em Proc. AAAI'1996}, vol.~96, pp.~226--231, 1996.

\bibitem{Gan2015ClusterLC}
M.~Gan, Z.~Xu, C.~F. Mecklenbr{\"a}uker, and T.~Zemen, ``Cluster lifetime
  characterization for vehicular communication channels,'' in {\em Proc.
  EuCAP'2015}, pp.~1--5, IEEE, 2015.

\bibitem{He2017AnAC}
R.~He, Q.~Li, B.~Ai, Y.~Geng, A.~Molisch, V.~Kristem, Z.~Zhong, and J.~Yu, ``An
  automatic clustering algorithm for multipath components based on
  kernel-power-density,'' {\em 2017 IEEE Wireless Commun. and Networking
  Conference (WCNC)}, pp.~1--6, 2017.

\bibitem{He2017AKA}
R.~He, Q.~Li, B.~Ai, Y.~Geng, A.~Molisch, V.~Kristem, Z.~Zhong, and J.~Yu, ``A
  kernel-power-density-based algorithm for channel multipath components
  clustering,'' {\em IEEE Trans. Wireless Commun.}, vol.~16, pp.~7138--7151,
  2017.

\bibitem{Huang2017ANP}
C.~Huang, R.~He, and Z.~Zhong, ``A novel power weighted multipath component
  tracking algorithm,'' in {\em Proc. URSI GASS'2017}, pp.~1--4, IEEE, 2017.

\bibitem{Huang2017ANT}
C.~Huang, R.~He, Z.~Zhong, Y.~ao~Geng, Q.~Li, and Z.~Zhong, ``A novel
  tracking-based multipath component clustering algorithm,'' {\em IEEE Antennas
  Wirel. \& Propag. Lett.}, vol.~16, pp.~2679--2683, 2017.

\bibitem{Wang2017AFO}
Q.~Wang, B.~Ai, R.~He, K.~Guan, Y.~Li, Z.~Zhong, and G.~Shi, ``A framework of
  automatic clustering and tracking for time-variant multipath components,''
  {\em IEEE Commun. Lett.}, vol.~21, pp.~953--956, 2017.

\bibitem{Cai2018HoughTransformBasedCI}
X.~Cai, B.~Peng, X.~Yin, and A.~P. Yuste, ``Hough-transform-based cluster
  identification and modeling for {V2V} channels based on measurements,'' {\em
  IEEE Trans. Veh. Technol.}, vol.~67, pp.~3838--3852, 2018.

\bibitem{Huang2019APB}
C.~Huang, R.~He, Z.~Zhong, B.~Ai, Y.~ao~Geng, Z.~Zhong, Q.~Li, K.~Haneda, and
  C.~Oestges, ``A power-angle-spectrum based clustering and tracking algorithm
  for time-varying radio channels,'' {\em IEEE Trans. Veh. Technol.}, vol.~68,
  pp.~291--305, 2019.

\bibitem{Huang2018ANT}
C.~Huang, R.~He, Z.~Zhong, B.~Ai, G.~Wang, Z.~Zhong, C.~Oestges, and K.~Haneda,
  ``A novel target recognition based radio channel clustering algorithm,'' in
  {\em Proc. WCSP'2018}, pp.~1--6, IEEE, 2018.

\bibitem{huang2018analysis}
C.~Huang, R.~He, Z.~Zhong, and Z.~Zhong, ``Analysis of edge detection for the
  clusters in radio propagation channel,'' in {\em Proc. IEEE
  APS/USNC-URSI'2018}, pp.~91--92, IEEE, 2018.

\bibitem{Wu2020ANP}
T.~Wu, X.~Yin, and J.~Lee, ``A novel power spectrum-based sequential tracker
  for time-variant radio propagation channel,'' {\em IEEE Access}, vol.~8,
  pp.~151267--151278, 2020.

\bibitem{Huang2020TrajectoryJointCA}
C.~Huang, A.~F. Molisch, Y.~Geng, R.~He, B.~Ai, and Z.~Zhong,
  ``Trajectory-joint clustering algorithm for time-varying channel modeling,''
  {\em IEEE Trans. Veh. Technol.}, vol.~69, pp.~1041--1045, 2020.

\bibitem{Huang2020ClusteringPE}
C.~Huang, R.~He, B.~Ai, M.~Yang, Y.-A. Geng, and Z.~Zhong, ``Clustering
  performance evaluation algorithm for vehicle-to-vehicle radio channels,'' in
  {\em Proc. EuCAP'2020}, pp.~1--4, IEEE, 2020.

\bibitem{toeltsch2002statistical}
M.~Toeltsch, J.~Laurila, K.~Kalliola, A.~F. Molisch, P.~Vainikainen, and
  E.~Bonek, ``Statistical characterization of urban spatial radio channels,''
  {\em IEEE J. Sel. Areas Commun.}, vol.~20, no.~3, pp.~539--549, 2002.

\bibitem{karedal2004uwb}
J.~Karedal, S.~Wyne, P.~Almers, F.~Tufvesson, and A.~F. Molisch, ``Uwb channel
  measurements in an industrial environment,'' in {\em Proc. IEEE
  GLOBECOM'2004}, vol.~6, pp.~3511--3516, IEEE, 2004.

\bibitem{Illingworth1988ASO}
J.~Illingworth and J.~Kittler, ``A survey of the hough transform,'' {\em
  Comput. Vis. Graph. Image Process.}, vol.~44, pp.~87--116, 1988.

\bibitem{Otsu1979ATS}
N.~Otsu, ``A threshold selection method from gray level histograms,'' {\em IEEE
  Trans. Syst. Man Cybern}, vol.~9, pp.~62--66, 1979.

\bibitem{Pedersen2016ClusteringBF}
A.~Rodriguez and A.~Laio, ``Clustering by fast search and find of density
  peaks,'' {\em science}, vol.~344, no.~6191, pp.~1492--1496, 2014.

\bibitem{Babtlett1948SmoothingPF}
M.~Babtlett, ``Smoothing periodograms from time-series with continuous
  spectra,'' {\em Nature}, vol.~161, pp.~686--687, 1948.

\bibitem{huang2020geometry}
C.~Huang, R.~Wang, P.~Tang, R.~He, B.~Ai, Z.~Zhong, C.~Oestges, and A.~F.
  Molisch, ``Geometry-cluster-based stochastic {MIMO} model for
  vehicle-to-vehicle communications in street canyon scenarios,'' {\em IEEE
  Trans. Wireless Commun.}, 2020.

\bibitem{yang2020measurements}
M.~Yang, B.~Ai, R.~He, G.~Wang, L.~Chen, X.~Li, C.~Huang, Z.~Ma, Z.~Zhong,
  J.~Wang, {\em et~al.}, ``Measurements and cluster-based modeling of
  vehicle-to-vehicle channels with large vehicle obstructions,'' {\em IEEE
  Trans. Wireless Commun.}, vol.~19, no.~9, pp.~5860--5874, 2020.

\bibitem{Aggarwal2003AFF}
C.~Aggarwal, J.~Han, J.~Wang, and P.~S. Yu, ``A framework for clustering
  evolving data streams,'' in {\em Proc. VLDB'2003}, 2003.

\bibitem{Maulik2002PerformanceEO}
U.~Maulik and S.~Bandyopadhyay, ``Performance evaluation of some clustering
  algorithms and validity indices,'' {\em IEEE Trans. Pattern Anal. Mach.
  Intell.}, vol.~24, pp.~1650--1654, 2002.

\bibitem{Pakhira2004ValidityIF}
M.~K. Pakhira, S.~Bandyopadhyay, and U.~Maulik, ``Validity index for crisp and
  fuzzy clusters,'' {\em Pattern Recognit.}, vol.~37, pp.~487--501, 2004.

\bibitem{Wang2007OnFC}
W.~Wang and Y.~Zhang, ``On fuzzy cluster validity indices,'' {\em Fuzzy Sets
  Syst.}, vol.~158, pp.~2095--2117, 2007.

\bibitem{Mota2013EstimationOT}
S.~Mota, F.~P{\'e}rez-Font{\'a}n, and A.~Rocha, ``Estimation of the number of
  clusters in multipath radio channel data sets,'' {\em IEEE Trans. Antennas
  Propag.}, vol.~61, pp.~2879--2883, 2013.

\bibitem{Brown1983IntroductionTR}
R.~G. Brown, {\em Introduction to random signal analysis and Kalman filtering}.
\newblock Wiley, 1983.

\bibitem{Bazaraa1977LinearPA}
M.~S. Bazaraa, J.~J. Jarvis, and H.~D. Sherali, ``Linear programming and
  network flows,'' John Wiley \& Sons, 1977.

\bibitem{Munkres1957AlgorithmsFT}
J.~Munkres, ``Algorithms for the assignment and transportation problems,'' {\em
  Journal of The Society for Industrial and Applied Mathematics}, vol.~10,
  pp.~196--210, 1957.

\bibitem{Hanpinitsak2017MultipathCA}
P.~Hanpinitsak, K.~Saito, J.~Takada, M.~Kim, and L.~Materum, ``Multipath
  clustering and cluster tracking for geometry-based stochastic channel
  modeling,'' {\em IEEE Trans. Antennas Propag.}, vol.~65, pp.~6015--6028,
  2017.

\bibitem{Hanpinitsak2019FrequencyCO}
P.~Hanpinitsak, K.~Saito, W.~Fan, J.~Hejselb{\ae}k, J.~Takada, and G.~Pedersen,
  ``Frequency characteristics of geometry-based clusters in indoor hall
  environment at {SHF} bands,'' {\em IEEE Access}, vol.~7, pp.~75420--75433,
  2019.

\bibitem{Materum2009MobileSS}
L.~Materum, J.~Takada, I.~Ida, and Y.~Oishi, ``Mobile station spatio-temporal
  multipath clustering of an estimated wideband {MIMO} double-directional
  channel of a small urban 4.5 {GHz} macrocell,'' {\em EURASIP Journal on
  Wireless Communications and Networking}, vol.~2009, pp.~1--16, 2009.

\bibitem{Wang2018RandomCN}
C.~Wang, J.~Zhang, and F.~Tufvesson, ``Random cluster number feature and
  cluster characteristics of indoor measurement at 28 {GHz},'' {\em IEEE
  Antennas Wirel. \& Propag. Lett.}, vol.~17, pp.~1881--1884, 2018.

\bibitem{Lai2019MethodologyFM}
C.~Lai, R.~Sun, C.~Gentile, P.~Papazian, J.~Wang, and J.~Senic, ``Methodology
  for multipath-component tracking in millimeter-wave channel modeling,'' {\em
  IEEE Trans. Antennas Propag.}, vol.~67, pp.~1826--1836, 2019.

\bibitem{Czink2006ANA}
N.~Czink, C.~Mecklenbraeuker, and G.~D. Galdo, ``A novel automatic cluster
  tracking algorithm,'' {\em 2006 IEEE 17th International Symposium on
  Personal, Indoor and Mobile Radio Communications}, pp.~1--5, 2006.

\bibitem{Karedal2009AGS}
J.~Karedal, F.~Tufvesson, N.~Czink, A.~Paier, C.~Dumard, T.~Zemen,
  C.~Mecklenbraeuker, and A.~Molisch, ``A geometry-based stochastic {MIMO}
  model for vehicle-to-vehicle communications,'' {\em IEEE Trans. Wireless
  Commun.}, vol.~8, 2009.

\bibitem{He2015ADW}
R.~He, O.~Renaudin, V.~Kolmonen, K.~Haneda, Z.~Zhong, B.~Ai, and C.~Oestges,
  ``A dynamic wideband directional channel model for vehicle-to-vehicle
  communications,'' {\em IEEE Trans. Ind. Electron.}, vol.~62, pp.~7870--7882,
  2015.

\bibitem{Mahler2017TrackingOW}
K.~Mahler, W.~Keusgen, F.~Tufvesson, T.~Zemen, and G.~Caire, ``Tracking of
  wideband multipath components in a vehicular communication scenario,'' {\em
  IEEE Trans. Veh. Technol.}, vol.~66, pp.~15--25, 2017.

\bibitem{Czink2007TrackingTC}
N.~Czink, R.~Tian, S.~Wyne, F.~Tufvesson, J.-P. Nuutinen, J.~Ylitalo, E.~Bonek,
  and A.~F. Molisch, ``Tracking time-variant cluster parameters in mimo channel
  measurements,'' in {\em Proc. CHINACOM'2007}, pp.~1147--1151, IEEE, 2007.

\bibitem{Gedschold2018TrackingBM}
J.~Gedschold, C.~Schneider, M.~K{\"a}ske, R.~Thom{\"a}, G.~D. Galdo, M.~Boban,
  and J.~Luo, ``Tracking based multipath clustering in
  vehicle-to-infrastructure channels,'' {\em 2018 IEEE 29th Annual
  International Symposium on Personal, Indoor and Mobile Radio Communications
  (PIMRC)}, pp.~1--5, 2018.

\bibitem{Arulampalam2002ATO}
M.~S. Arulampalam, S.~Maskell, N.~Gordon, and T.~Clapp, ``A tutorial on
  particle filters for online nonlinear/non-gaussian bayesian tracking,'' {\em
  IEEE Trans. Signal Process.}, vol.~50, pp.~174--188, 2002.

\bibitem{Yin2008TrackingOT}
X.~Yin, G.~Steinbock, G.~E. Kirkelund, T.~Pedersen, P.~Blattnig, A.~Jaquier,
  and B.~H. Fleury, ``Tracking of time-variant radio propagation paths using
  particle filtering,'' in {\em Proc. IEEE ICC'2008}, pp.~920--924, IEEE, 2008.

\bibitem{Butler1961BeamformingMS}
J.~Butler and R.~Lowe, ``Beam-forming matrix simplifies design of
  electronically scanned antennas,'' {\em Electronic Design}, vol.~9,
  pp.~170--173, 1961.

\bibitem{wilson1992electrical}
G.~Wilson, ``Electrical downtilt through beam-steering versus mechanical
  downtilt (base station antennas),'' in {\em Proc. Vehicular Technology
  Society 42nd VTS Conference-Frontiers of Technology'1992}, pp.~1--4, IEEE,
  1992.

\bibitem{mesleh2008spatial}
R.~Y. Mesleh, H.~Haas, S.~Sinanovic, C.~W. Ahn, and S.~Yun, ``Spatial
  modulation,'' {\em IEEE Trans. Veh. Technol.}, vol.~57, no.~4,
  pp.~2228--2241, 2008.

\bibitem{klautau20185g}
A.~Klautau, P.~Batista, N.~Gonz{\'a}lez-Prelcic, Y.~Wang, and R.~W. Heath,
  ``{5G} {MIMO} data for machine learning: Application to beam-selection using
  deep learning,'' in {\em Proc. ITA'2018}, pp.~1--9, IEEE, 2018.

\bibitem{sim2020deep}
M.~S. Sim, Y.-G. Lim, S.~H. Park, L.~Dai, and C.-B. Chae, ``Deep learning-based
  mmwave beam selection for {5G NR/6G} with sub-6 {GHz} channel information:
  Algorithms and prototype validation,'' {\em IEEE Access}, vol.~8,
  pp.~51634--51646, 2020.

\bibitem{8734054}
Y.~{Wang}, A.~{Klautau}, M.~{Ribero}, A.~C.~K. {Soong}, and R.~W. {Heath},
  ``Mmwave vehicular beam selection with situational awareness using machine
  learning,'' {\em IEEE Access}, vol.~7, pp.~87479--87493, 2019.

\bibitem{8662770}
V.~{Va}, T.~{Shimizu}, G.~{Bansal}, and R.~W. {Heath}, ``Online learning for
  position-aided millimeter wave beam training,'' {\em IEEE Access}, vol.~7,
  pp.~30507--30526, 2019.

\bibitem{8503086}
A.~{Klautau}, P.~{Batista}, N.~{Gonz\'alez-Prelcic}, Y.~{Wang}, and R.~W.
  {Heath}, ``{5G} {MIMO} data for machine learning: Application to
  beam-selection using deep learning,'' in {\em Proc. ITA'2018}, pp.~1--9,
  2018.

\bibitem{8642397}
A.~{Klautau}, N.~{González-Prelcic}, and R.~W. {Heath}, ``{LIDAR} data for
  deep learning-based mmwave beam-selection,'' {\em IEEE Wireless Commun.
  Letters}, vol.~8, no.~3, pp.~909--912, 2019.

\bibitem{guo2013spectral}
W.~Guo, S.~Wang, Y.~Wu, J.~Rigelsford, X.~Chu, and T.~O'Farrell, ``Spectral-and
  energy-efficient antenna tilting in a hetnet using reinforcement learning,''
  in {\em Proc. IEEE WCNC'2013}, pp.~767--772, IEEE, 2013.

\bibitem{dandanov2017dynamic}
N.~Dandanov, H.~Al-Shatri, A.~Klein, and V.~Poulkov, ``Dynamic
  self-optimization of the antenna tilt for best trade-off between coverage and
  capacity in mobile networks,'' {\em Wireless Personal Communications},
  vol.~92, no.~1, pp.~251--278, 2017.

\bibitem{parera2018transferring}
C.~Parera, A.~E. Redondi, M.~Cesana, Q.~Liao, L.~Ewe, and C.~Tatino,
  ``Transferring knowledge for tilt-dependent radio map prediction,'' in {\em
  Proc. IEEE WCNC'2018}, pp.~1--6, IEEE, 2018.

\bibitem{capon1969high}
J.~Capon, ``High-resolution frequency-wavenumber spectrum analysis,'' {\em
  Proc. IEEE}, vol.~57, no.~8, pp.~1408--1418, 1969.

\bibitem{long2018data}
Y.~Long, Z.~Chen, J.~Fang, and C.~Tellambura, ``Data-driven-based analog beam
  selection for hybrid beamforming under mm-wave channels,'' {\em IEEE J. Sel.
  Top. Signal Process.}, vol.~12, no.~2, pp.~340--352, 2018.

\bibitem{lecci2020machine}
M.~Lecci, P.~Testolina, M.~Rebato, A.~Testolin, and M.~Zorzi, ``Machine
  learning-aided design of thinned antenna arrays for optimized network level
  performance,'' in {\em Proc. EuCAP'2020}, pp.~1--5, IEEE, 2020.

\bibitem{guo2017deep}
X.~Guo, X.~Liu, E.~Zhu, and J.~Yin, ``Deep clustering with convolutional
  autoencoders,'' in {\em Proc. ICONIP'2017}, pp.~373--382, Springer, 2017.

\bibitem{ghasedi2017deep}
K.~Ghasedi~Dizaji, A.~Herandi, C.~Deng, W.~Cai, and H.~Huang, ``Deep clustering
  via joint convolutional autoencoder embedding and relative entropy
  minimization,'' in {\em Proc. CVPR'2017}, pp.~5736--5745, 2017.

\bibitem{huang2014deep}
P.~Huang, Y.~Huang, W.~Wang, and L.~Wang, ``Deep embedding network for
  clustering,'' in {\em Proc. ICPR'2014}, pp.~1532--1537, IEEE, 2014.

\bibitem{glazunov2009spherical}
A.~A. Glazunov, M.~Gustafsson, A.~F. Molisch, F.~Tufvesson, and G.~Kristensson,
  ``Spherical vector wave expansion of gaussian electromagnetic fields for
  antenna-channel interaction analysis,'' {\em IEEE Trans. Antennas Propag.},
  vol.~57, no.~7, pp.~2055--2067, 2009.

\bibitem{evans2010near}
D.~N. Evans and M.~A. Jensen, ``Near-optimal radiation patterns for antenna
  diversity,'' {\em IEEE Trans. Antennas Propag.}, vol.~58, no.~11,
  pp.~3765--3769, 2010.

\end{thebibliography}

\begin{IEEEbiography}[{\includegraphics[width=1in,height=1.25in,clip,keepaspectratio]{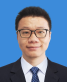}}]
{Chen Huang} (S'17-M'21) received the Ph.D. degrees from Beijing Jiaotong University, Beijing, China, in 2021. He is currently a research fellow in the Pervasive Communication Research Center, Purple Mountain Laboratories, Nanjing, 211111, China, and also a Post Doc in the National Mobile Communications Research Laboratory, School of Information Science and Engineering, Southeast University, Nanjing, 210096, China. His research interests are in machine-learning-based channel modeling.
\end{IEEEbiography}
\vspace{-1 cm}

\begin{IEEEbiography}[{\includegraphics[width=1in,height=1.25in,clip,keepaspectratio]{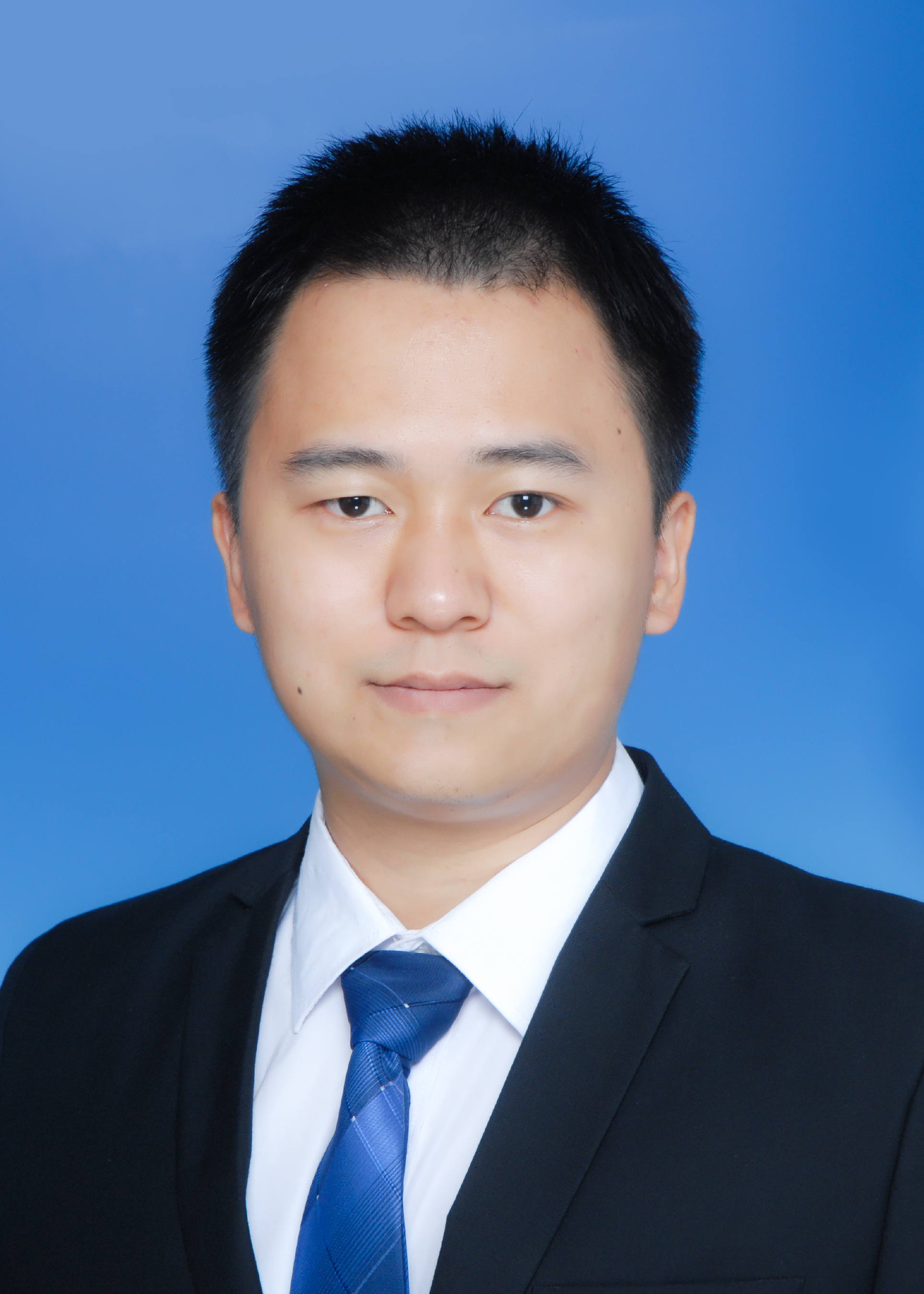}}]
{Ruisi He} (S'11-M'13-SM'17) received the B.E. and Ph.D. degrees from Beijing Jiaotong University, Beijing, China, in 2009 and 2015, respectively. Since 2015, Dr. He has been with the State Key Laboratory of Rail Traffic Control and Safety, BJTU, where he has been a Full Professor since 2018. Dr. He has been a Visiting Scholar in Georgia Institute of Technology, USA, University of Southern California, USA, and Universit\'e Catholique de Louvain, Belgium. His research interests include wireless propagation channels, railway and vehicular communications, 5G and 6G communications.
\end{IEEEbiography}
\vspace{-1 cm}

\begin{IEEEbiography}[{\includegraphics[width=1in,height=1.25in,clip,keepaspectratio]{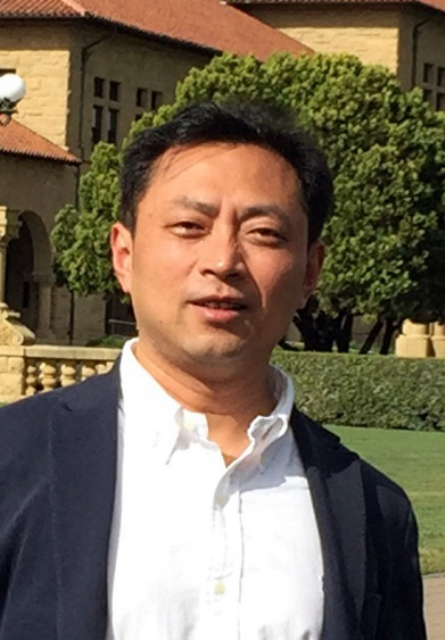}}]
{Bo Ai} (M'00-SM'10) is currently a Professor and an Advisor of Ph.D. candidates with Beijing Jiaotong University, Beijing, where he is also the Deputy Director of the State Key Laboratory of Rail Traffic Control and Safety. He is also currently with the Engineering College, Armed Police Force, Xian.  His interests include the research and applications of orthogonal frequency-division multiplexing techniques, high-power amplifier linearization techniques, radio propagation and channel modeling, global systems for mobile communications for railway systems, and long-term evolution for railway systems.
\end{IEEEbiography}
\vspace{-1 cm}

\begin{IEEEbiography}[{\includegraphics[width=1in,height=1.25in,clip,keepaspectratio]{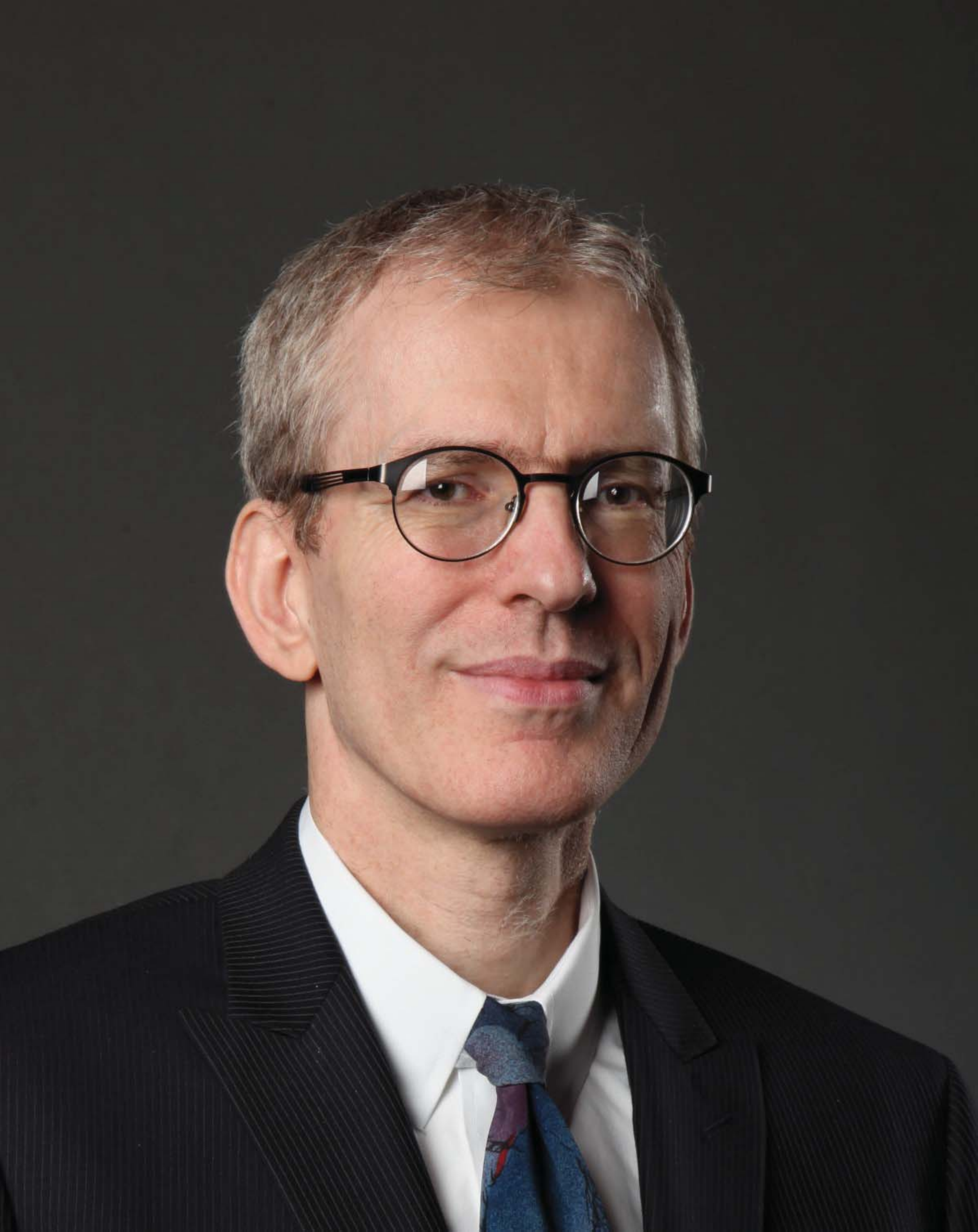}}]
{Andreas F. Molisch} (S'89-M'95-SM'00-F'05)  is the Solomon Golomb Andrew and Erna Viterbi Chair Professor at the University of Southern California. He was previously at TU Vienna, AT\&T (Bell) Labs, Lund University, and Mitsubishi Electric Research Labs. His research interests are in wireless communications, with emphasis on propagation channels, multiantenna systems, ultrawideband systems, and localization. He is a Fellow
of NAI, AAAS, and IET, a member of the Austrian Academy of Sciences, and a recipient of numerous awards.
\end{IEEEbiography}
\vspace{-1 cm}

\begin{IEEEbiography}[{\includegraphics[width=1in,height=1.25in,clip,keepaspectratio]{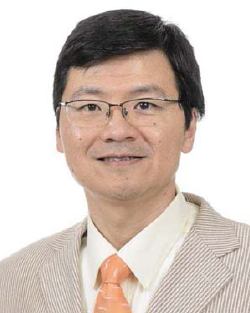}}]
{Buon Kiong Lau}(Senior Member, IEEE) received the B.E. degree (Hons.) in electrical engineering from the University of Western Australia, Perth, WA, Australia, in 1998, and the Ph.D. degree in electrical engineering from the Curtin University of Technology, Perth, in 2003. Since 2004, he has been with the Department of Electrical and Information Technology, Lund University, where he is currently a Professor. His primary research interests are in various aspects of multiple antenna systems, particularly the interplay between antennas, propagation channels, and signal processing.
\end{IEEEbiography}
\vspace{-1 cm}

\begin{IEEEbiography}[{\includegraphics[width=1in,height=1.25in,clip,keepaspectratio]{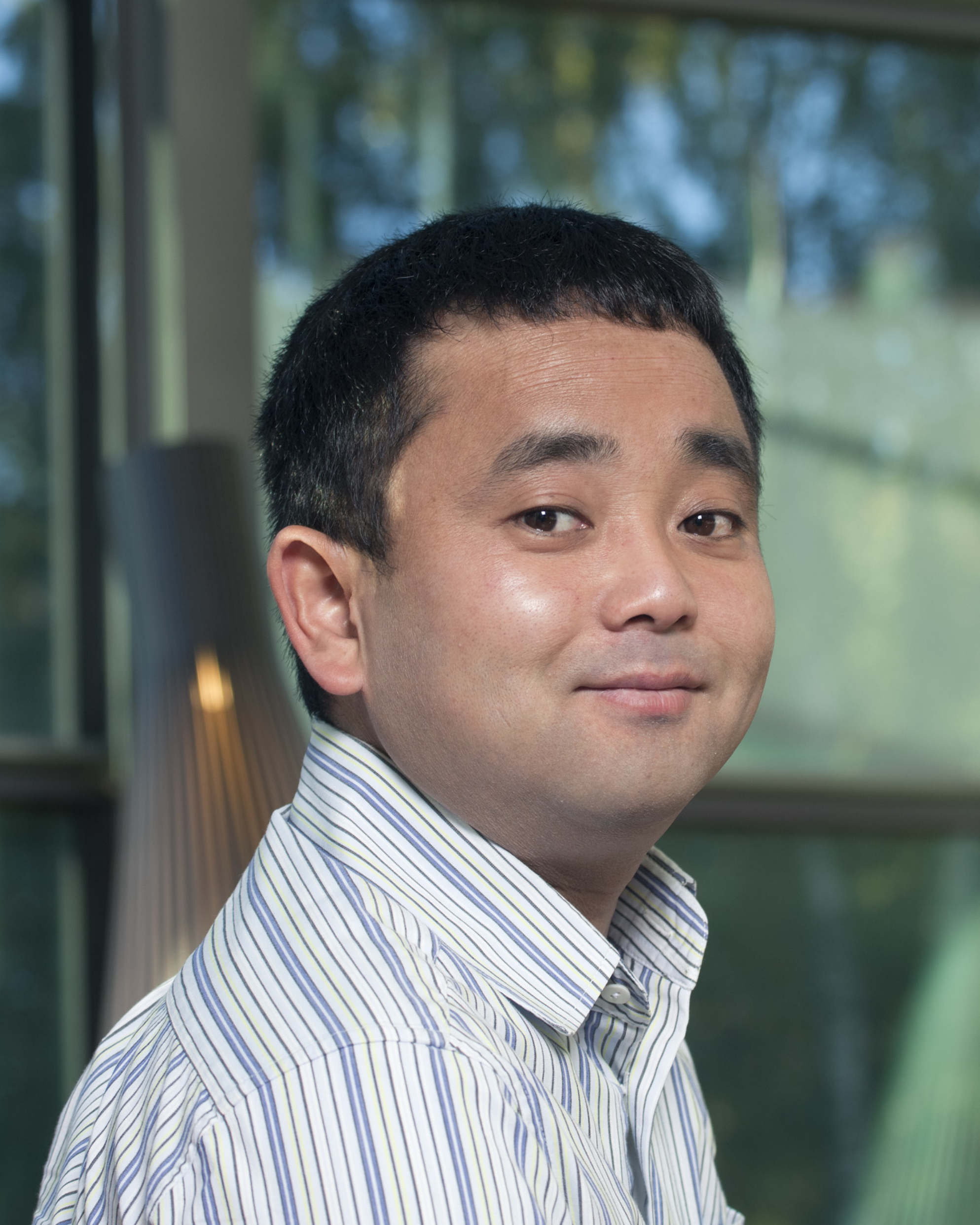}}]
{Katsuyuki Haneda} is an associate professor in the Aalto University, Finland. He has been an associate editor of the IEEE Transactions on Antennas and Propagation for 2012-2016, and of an editor of the IEEE Transactions on Wireless Communications for 2013-2018. His current research activity includes radio frequency instrumentation, measurements and modeling, millimeter-wave radios, in-band full-duplex radio technology and radio applications in medical and healthcare scenarios.
\end{IEEEbiography}
\vspace{-1 cm}

\begin{IEEEbiography}[{\includegraphics[width=1in,height=1.25in,clip,keepaspectratio]{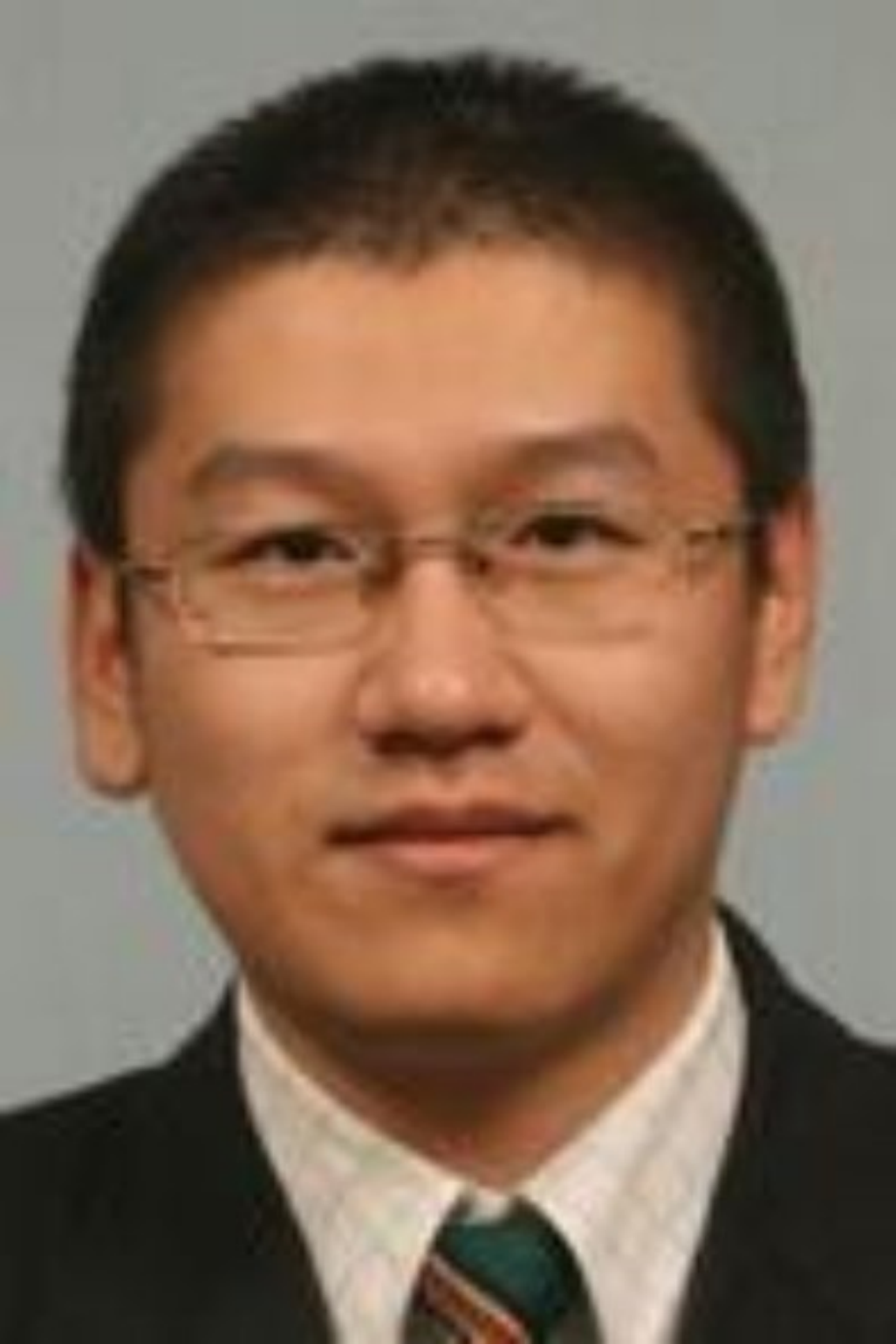}}]
	{Bo Liu}(M'15-SM'17) received the B.S. degree from Tsinghua University, Beijing, China, in 2008, and the Ph.D. degree from the University of Leuven (KU Leuven), Leuven, Belgium, in 2012. He is currently a Senior Lecturer (Associate Professor) with the University of Glasgow, Glasgow, U.K. He is also a Senior Honorary Fellow with the University of Birmingham, Birmingham, U.K.. His research interests lie in artificial intelligence-driven design methodologies of analog/RF integrated circuits, microwave devices, MEMS, evolutionary computation, and machine learning.
\end{IEEEbiography}
\vspace{-1 cm}

\begin{IEEEbiography}[{\includegraphics[width=1in,height=1.25in,clip,keepaspectratio]{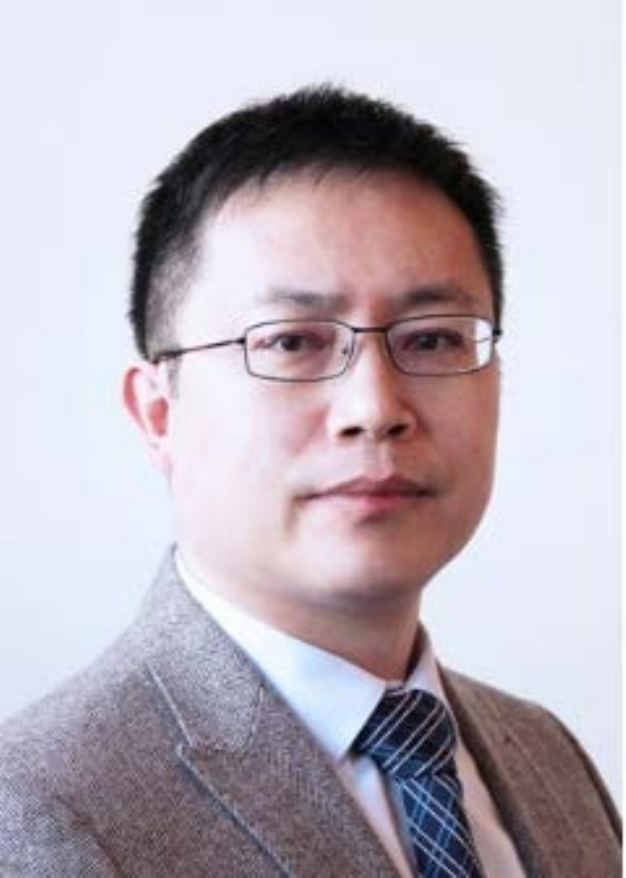}}]
	{Cheng-Xiang Wang}(S'01-M'05-SM'08-F'17) received the BSc and MEng degrees in Communication and Information Systems from Shandong University, China, in 1997 and 2000, respectively, and the PhD degree in Wireless Communications from Aalborg University, Denmark, in 2004. In 2018, he joined the National Mobile Communications Research Laboratory, Southeast University, China, as a Professor. He is also a parttime professor with the Purple Mountain Laboratories, Nanjing, China. His current research interests include wireless channel measurements and modeling, B5G wireless communication networks, and applying artificial intelligence to wireless networks.
 \end{IEEEbiography}
 \vspace{-1 cm}

\begin{IEEEbiography}[{\includegraphics[width=1in,height=1.25in,clip,keepaspectratio]{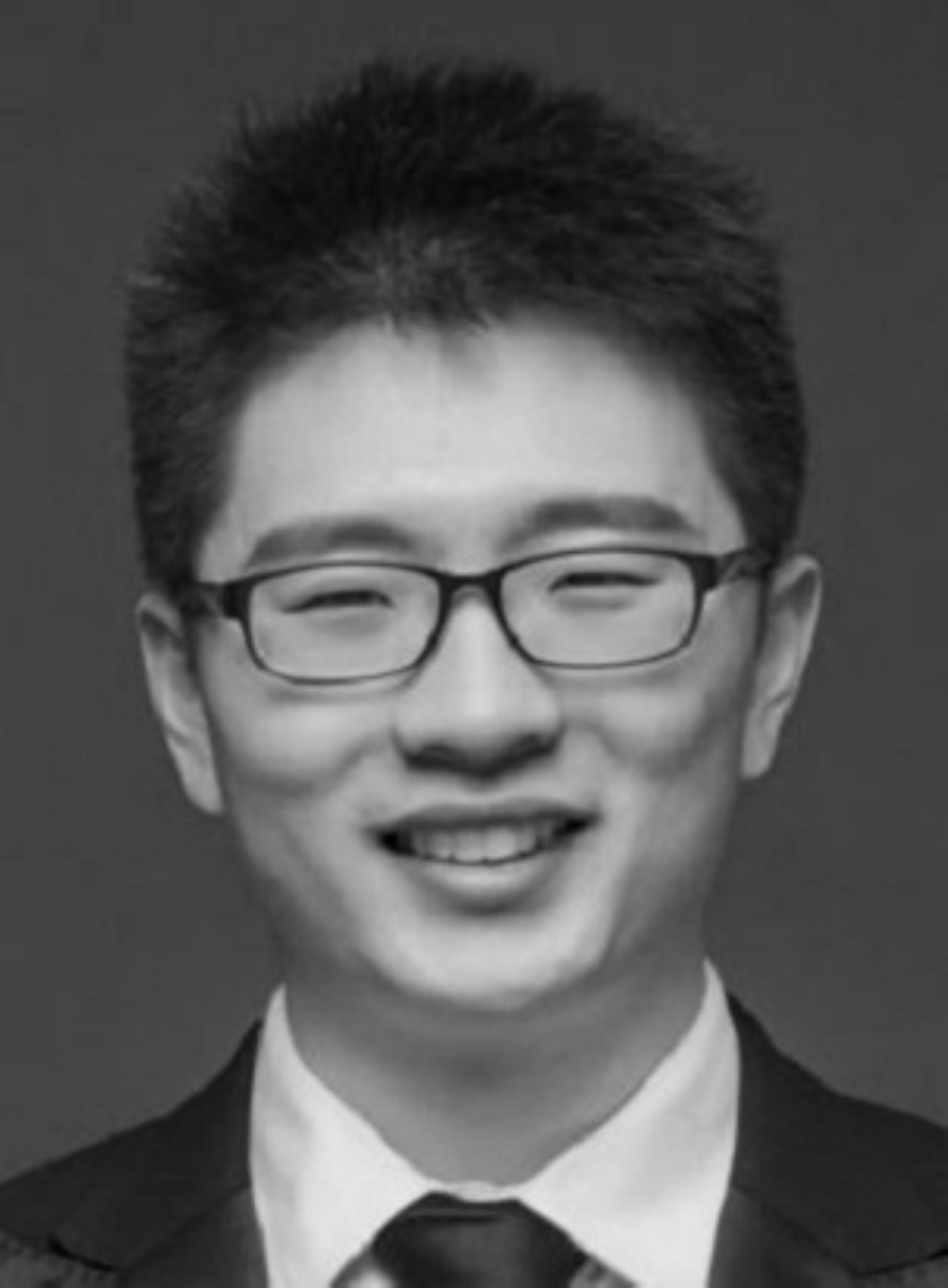}}]
{Mi Yang} (S'17-M'21) received the M.S. and Ph.D. degrees from Beijing Jiaotong University (BJTU), Beijing, China, in 2017 and 2021, respectively. He is currently a associate professor with the State Key Laboratory of Rail Traffic Control and Safety, Beijing Jiaotong University. His research interests are focused on wireless propagation channels, vehicle-to-everything (V2X) communications and 5G/B5G communications.
\end{IEEEbiography}
\vspace{-1 cm}

\begin{IEEEbiography}[{\includegraphics[width=1in,height=1.25in,clip,keepaspectratio]{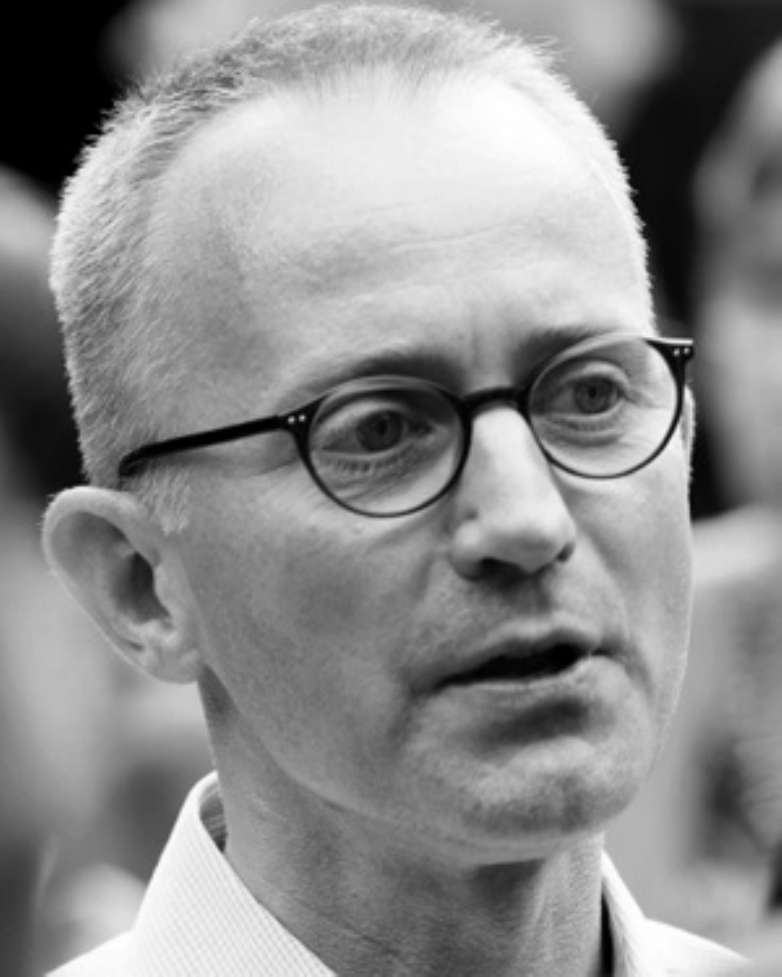}}]
	{Claude Oestges}(Fellow, IEEE) received the M.Sc and Ph.D. degrees in Electrical Engineering from the  Universit\'e Catholique de Louvain (UCLouvain), \'Louvain-la-Neuve, Belgium, in 1996 and 2000, respectively. In January 2001, he joined the Smart Antennas Research Group (Information Systems Laboratory), Stanford University, Stanford, CA, USA, as a Postdoctoral Scholar. From January 2002 to September 2005, he was associated with the Microwave Laboratory, UCLouvain, as a Postdoctoral Fellow of the Belgian Fonds de la Recherche Scientifique (FRS-FNRS). He is currently a Full Professor with the Electrical Engineering Department, Institute for Information and Communication Technologies, Electronics and Applied Mathematics (ICTEAM), UCLouvain. 
\end{IEEEbiography}
\vspace{-1 cm}

\begin{IEEEbiography}[{\includegraphics[width=1in,height=1.25in,clip,keepaspectratio]{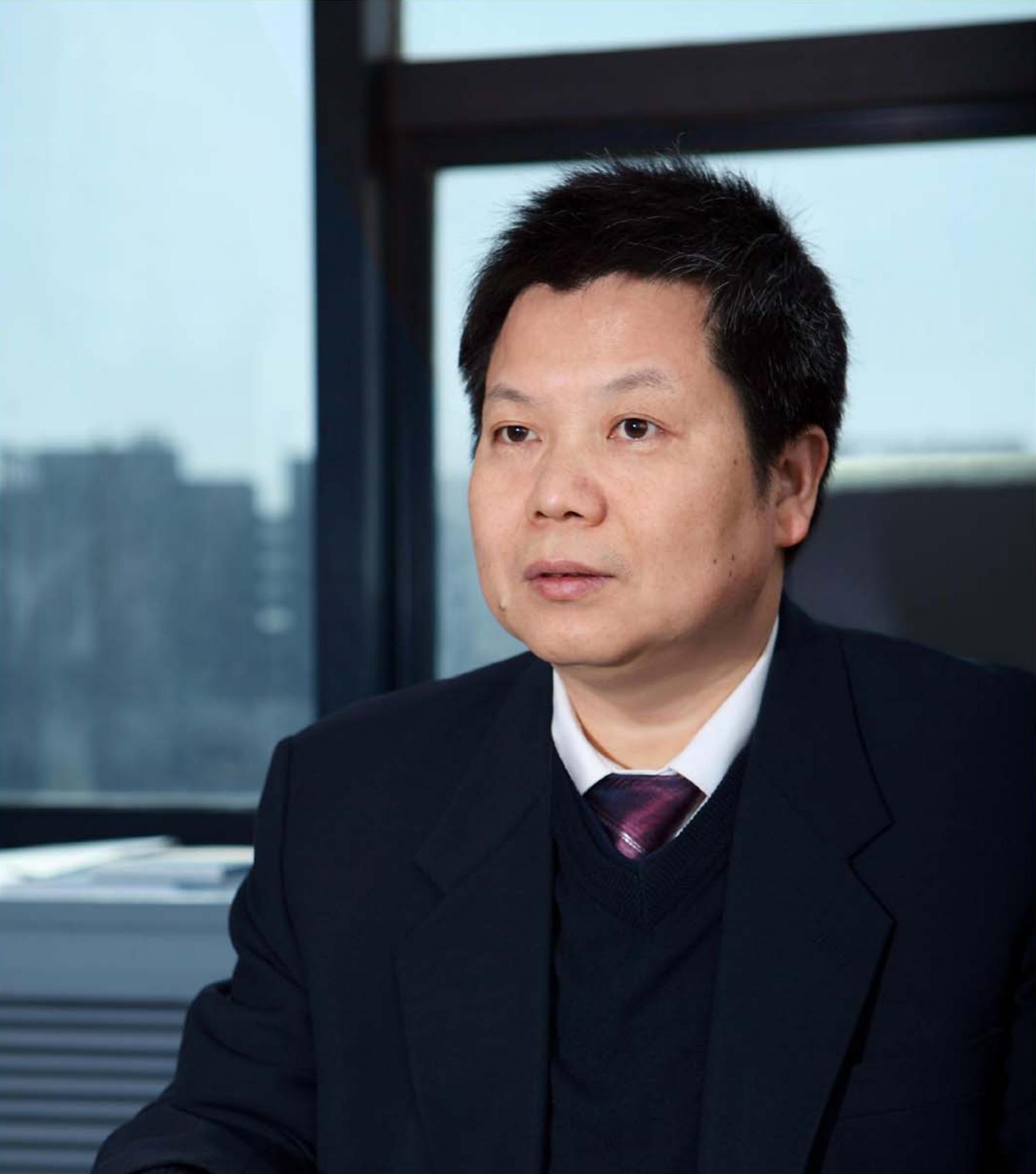}}]
	{Zhangdui Zhong} (SM'16) received the B.E. and M.S. degrees from Beijing Jiaotong University, Beijing, China, in 1983 and 1988, respectively. He is a Professor and Advisor of Ph.D. candidates with Beijing Jiaotong University, Beijing, China. He is currently a Director of the School of Computer and Information Technology and a Chief Scientist of State Key Laboratory of Rail Traffic Control and Safety, Beijing Jiaotong University. He is an Executive Council Member of Radio Association of China, Beijing, and a Deputy Director of Radio Association, Beijing. His interests include wireless communications for railways, control theory and techniques for railways, and GSM-R systems. 
\end{IEEEbiography}

%
%
%

%
%
%
%
%




\end{document}